\documentclass[intlimits,twoside,a4paper]{article}

\usepackage[cp1251]{inputenc}
\usepackage{cmpj3}

\usepackage{enumitem}

\chardef\bslash=`\\ 

\newcommand{\be}{\begin{equation}}
\newcommand{\ee}{\end{equation}}
\newcommand{\bea}{\begin{eqnarray}}
\newcommand{\eea}{\end{eqnarray}}

\articletype{A part of the Special collection to the 100th anniversary of the birth of Ihor Yukhnovskii}

\issue{2025}{28}{2}{23101}
\doinumber{10.5488/CMP.28.23101}

\title[Statistical theory of charged particle systems]%
{Statistical theory of charged particle systems including triple bound states --- and the Collaboration Lviv--Rostock}

\author[W. Ebeling]{W. Ebeling\orcid{0000-0003-0740-3016}\thanks{Email: \email{ebeling@physik.hu-berlin.de}.}}

\address{
Institute of Physics, Humboldt University, Berlin, Germany
}

\date{Received March 09, 2025, in final form April 17, 2025}

\begin{document}
	\maketitle
\begin{abstract}
Honoring the hundredth anniversary of the birthday of Ihor R. Yuknovskii we analyze new developments in the statistical
thermodynamics of Coulomb systems. The basic idea of this work is to demonstrate that the exponential potential used in the first papers of
Yukhnovskii is an appropriate reference system for a description of classical and quantum charged particle systems. We briefly discuss the collaboration between the groups of Ihor R. Yuknovskii in Lviv and G\"unter Kelbg in Rostock and analyze several approaches based on pair correlation functions and cluster expansion in the classical as well as in the quantum case.
Finally, we discuss the progress in the statistical description of bound states of three particles as in helium plasmas and in MgCl$_2$-solutions in the classical case and present new results regarding the influence of three-particle bound states. In particular,  we give new expressions for the cluster integrals and the mass action functions of helium atoms and ionic triple associates
as well as for the equation of state (EoS). 
\end{abstract}

\hspace{8em}{{Dedicated to the 100th birthday of Ihor R. Yukhnovskii  (01.09.1925--26.03.2024)}} 
\vspace{1ex}


\section{Ihor Yukhnovskii and the statistical theory of systems of charged particles}
\label{intro}
The ``Statistical theory of equilibrium systems of charged particles'' is one of the main topics of the work of Ihor R. Yuknovskii and is also the title of his habilitation thesis, which he defended in 1965 at the Taras Shevchenko University in Kyiv \cite{Yukhn02}. We want to show here that besides his big influence on the development of science and society in Ukraine, Ihor Yukhnovskii had a large international impact, in particular on a scientific school at Rostock University. At this University near the coast of the Baltic Sea, the research on charged particles also had a high priority due to the pioneering work of Prof. Hans Falkenhagen (1895--1971) and G\"unter Kelbg (1922--1988). 
At this time the basic ideas in this field were developed in the school of Peter Debye, the teacher of Hans Falkenhagen. The ``Debye--H\"uckel-- Onsager--Falkenhagen theory'' was quite successful in interpreting data. However, this theory was still very much based on intuitive methods  \cite{Falkenhagen,Falkenhagen_a,Barthel}.  
 Only in the 40--50th of the 20th century the statistical  theory of many-particle effects in Coulombic systems was systematically developed using distribution functions by Nikolai~N.~Bogoliubov \cite{Bogoliubov,Bogoliubov_a} and in a different line based on a graph technique by Mayer,  Haga, and  Poirier, and, most systematically, by Harold~Friedman~\cite{Friedman}. 
The third line of systematic statistical approach to Coulomb systems was based on collective variables and is due to Ihor R. Yukhnovskii in Lviv and G\"unter Kelbg
in Rostock.

The idea to model ions in solution interacting as effective charged spheres was proposed by Peter~Debye, Erich H\"uckel and Lars Onsager and since then is a standard tool of electrolyte theory~\cite{Falkenhagen,Falkenhagen_a,Barthel,Friedman,GlaYuk52}. The common idea of the works of G\"unter Kelbg and Ihor Yukhnovslii was to use potentials possessing a 
Fourier transform and describes this way the screening effects in a most elegant way. The method also allows one to introduce collective variables and to develop a quite effective mathematical technique following the methods by Bohm, Pines and Zubarev \cite{Yukhn54,Yukhn58,Yukhn80,Kelbg59,Kelbg59a,Kelbg62,Kelbg62a,Kelbg63,KelbgHoff64,BoEbal22}.

 G\"unter Kelbg's and Ihor R. Yukhnovskii's students Norbert Albehrendt, Hartmut Hetzheim and Gri\-gori~Bigun continued this work
\cite{EbKe66,EbKe66a,Albehrendt,EbKr71,Czerwon}.
In particular, the late Hartmut  Krienke (1943--2023), developed their work for the classical case with many fruitful applications to electrolytes 
\cite{EbKr79}. He also continued  collaboration between the groups in Rostock and in Lviv and researched together with Myroslav Holovko, e.g., generalized virial expansions and solvation effects \cite{GoKr89,Krienke2013}. 
Further, he developed and analyzed several more general models of effective interactions for electrolytes with non-additive radii \cite{EbFeKr21,EbFeKr21a,EFKS77,EFKS77_a,EbFeSa79,EbKr2023}.
In some common work with Hartmut Krienke we discussed several
more general models of effective interactions for electrolytes. For example, we discussed in detail the model of charged spheres
with non-additive radii \cite{EbFeKr21,EbFeKr21a}.

In their first approach to electrolytes, Kelbg and Yukhnovskii used the strict expansions regarding the plasma parameter which were 
developed in 1946 by Nikolai N. Bogoliubov, solving the so-called BBGKY hierarchy with systematic expansion methods
\cite{Bogoliubov,Bogoliubov_a} and developed applications to the individual macroscopic properties of ionic species. Individual activity coefficients and individual electrical conductivities of ions are of high relevance for many problems connected with electrolytes. 
Some basic elements of an extension from Coulomb systems to the more general exponential potentials were given in \cite{EbKr79}. A detailed description of the Glauberman--Yukhnovskii--Kelbg theory including an extensive representation of many successful applications to a comparison of the experimental data is given in special chapter in the monograph of Falkenhagen \cite{Falkenhagen,Falkenhagen_a}. Here one finds 20 pages of a comprehensive 
presentation of the Glauberman--Yukhnovskii--Kelbg theory. 
Despite much success in the description of data, the work on the exponential potential should still be continued \cite{EbHoKuYu23}. 
In his heritage, Hartmut Krienke  left some notes with the plans for proceeding with the exponential potential. 
The general idea was to consider a species-independent simple exponential potential as some reference system and treat the short range perturbation by some type of generalized cluster expansion, similar to the one devised in~\cite{EbKr71,EbKr79,GoKr89}.

This work led, in particular, to a summary of the existing theoretical knowledge on the activity coefficients of electrolytes and extended also the theory 
to mean-spherical approximations and led to applications to transport properties. Special interest was devoted
to the consequences of charge asymmetry and of higher charging for the individual ionic properties. This way new information was obtained about the hydration sphere which strongly depends on the charges. The unfinished plan of Hartmut Krienke was to proceed along the same line for species-dependent exponential interactions \cite{EbHoKuYu23}.  The collaboration between the groups of Kelbg in Rostock and Yukhnovskii in Lviv 
has also been continued in the field of quantum statistics of Coulomb systems \cite{Albehrendt,EbHoKe67,EbHoKe67a,EbHoKe67b,KelbgEb70,KelbgEb70a}. Some results of this collaboration, as well as recent developments, will be discussed here.

\section{Classical systems with exponential interactions}
\subsection{Basic assumptions and model}
In the 1950's, Ihor R. Yukhnovskii developed together with his adviser Glauberman \cite{Yukhn02,GlaYuk52} a theory of electrolytes based on the method of Bogoliubov \cite{Bogoliubov,Bogoliubov_a}. This theory was later extended to the range of higher concentrations in many subsequent works, in particular by using collective coordinates  \cite{Yukhn02,Yukhn80}. For the description of the forces between charged particles Yukhnovskii and Glauberman used the effective potential, that had already been proposed  in 1927 by Kramers in quantum theory \cite{Kramers} and in 1934 by Hellman in quantum chemistry \cite{Hellmann,Hellmann_a,Hellmann_b}. We refer to this class of potentials as exponential potentials and consider first the simplest form 
\bea
V_{ij}(r) = \frac{ Z_i Z_j  e^2}{4 \piup \epsilon_0 \epsilon_r r} [1 - \exp (-\alpha r)],
\label{eq_1}
\eea
where $Z_i$ are the charge numbers. The exponential potential introduced by Kramers and Hellmann has a finite height at $r = 0$. 
Glauberman and Yukhnovskii \cite{GlaYuk52,Yukhn80} used this potential  for 
solving equilibrium problems of electrolyte theory and Kelbg 
developed successful applications to transport problems, in particular, applications to conductance and viscosity of electrolytes \cite{Falkenhagen,Falkenhagen_a,Kelbg59,Kelbg59a}. Kelbg extended the potential to the quantum-statistical case and studied the applications to plasmas.
 
\begin{figure}[htbp]
\begin{center}
\includegraphics[scale=4.93]{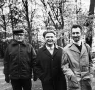}
\includegraphics[scale=0.365]{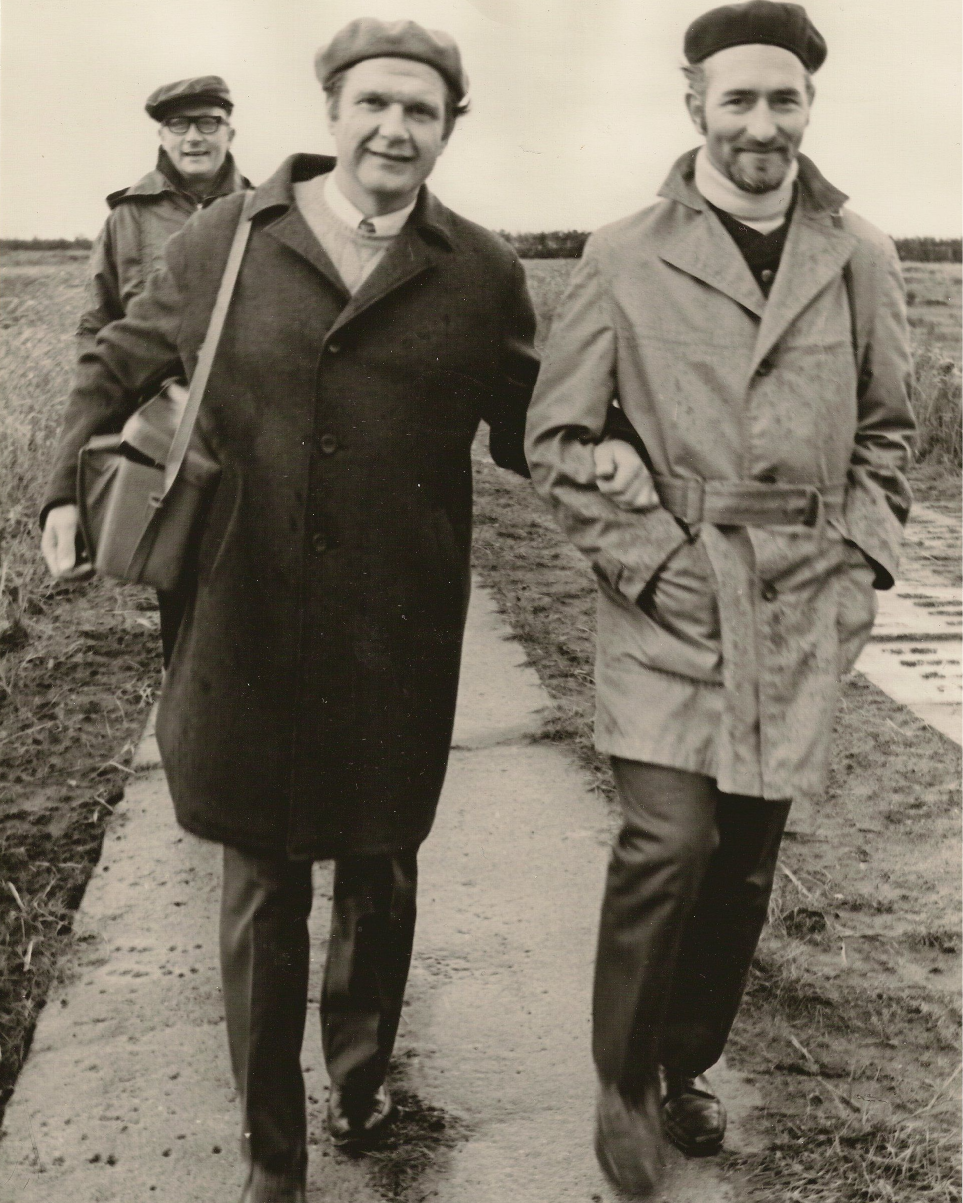}
\caption{Ihor Yukhnovskii was fond of nature and liked to walk or to jog in woods. Here we see him on a visit to University of Rostock in 1972 walking in the forest near the Baltic Sea and walking along the beach with his colleagues G\"unter Kelbg (left-hand) and Werner Ebeling (right-hand) (Fotos: Dorit Hagen).}
\label{Yukhn}
\end{center}
\end{figure}

These calculations were based on the general concepts of the Mc Millan--Mayer theory of solutions \cite{Friedman}.
We used approximations of the nonlinear Debye--H\"uckel and  Onsager--Fuoss type, then calculated pressure, osmotic coefficient, electrical and free energy \cite{EbKr71}. 
By Debye-charging we obtained the Helmholtz free energy, and the individual activity coefficients
of electrolytes with concentration $n_0$, the ion densities $n_i = \nu_i n_0$, the ion charges $e_i = Z_i e$, and the ionic strength $I = n_0 \sum_i \nu_i z_i^2 /2$ .
The potentials of the average force between the ions $i$ and $j$, $\psi_{ij}$, are usually separated into the long-range Coulomb potential
and a short-range potential $V_{ij}'(r)$. Here we proceed by including an additional exponential part
and a short-range part
\bea
 \psi_{ab} =\frac{ Z_a Z_b  \ell}{ r} k_{\rm B} T  [1 - \exp(-\alpha r)] + V_{ab}'(r); \qquad \ell = \frac{e^2}{4 \piup \epsilon_0 \epsilon_r k_{\rm B} T}.
\eea
Here, $\ell$ is the Coulomb length and $\epsilon_r (T, p)$ is the relative dielectric constant of pure water or solvent.
The length $\ell$ is also called Landau length or (double) Bjerrum length and is a function of
temperature and pressure. Here, most calculations are for the temperature $ T = 298.15$~K (i.e., $25^{\circ}$C)
and $\epsilon_r = 78.36$, the Coulomb length is then $\ell = 715.4$~pm.
The simplest possibility in our framework is to identify the long range part $V_{ij}$ with the exact Coulomb potential and 
to model the short-range repulsion by hard spheres with adaptable constact distances \cite{EbKr71,EbFeKr21,EbFeKr21a}.
In order to better describe the physical situation around an ion, in the older work square well and step
potentials were introduced \cite{Falkenhagen,Falkenhagen_a,Barthel,EFKS77,EFKS77_a,EbFeSa79}. The hard-core distances were fixed for this model as the sum of Pauling radii.
The remaining free potential parameters for the square-well potential were fitted to thermodynamic data, and the parameters of the step potential $h_{ij}$ were fitted to both thermodynamic data and to the conductance
data  for $c \leqslant 0.3$~mol/l for alkali halides.  The only fixed parameter was for iodid ions $d_{JJ}= 0$ which corresponds to
$h_{JJ} = 1.28$ (in units $k_{\rm B} T$). Physically, this corresponds to the a priori assumption that the iodide-ion is not hydrated. 

Replacing here the pure Coulomb potential by an exponential potential allows one  to take into account several effects, such as 
 strong deviations at small distances from Coulomb's law due the hydration in solutions and
due to Heisenberg uncertainty effects in quantum plasmas.  As shown in particular by Kelbg, the first order theory for the exponential potential is aso analytically solvable as for the pure Coulomb potential but leads already to an enormous improvement.  

Note that more appropriate for the description of the situation in the solvation region and near it may be the 
more flexible versions of the exponential potential  \cite{EbKr79,EbFeSa79}:
\bea
V_{ij}(r) =  Z_i Z_j k_{\rm B} T (\ell / r) [1 - \exp(-\alpha_{ij} r)]
\eea
or the equivalent ``ansatz''
\bea
V_{ij}(r) =  Z_i Z_j k_{\rm B} T (\ell / r) [1 - B_{ij} \exp(-\alpha r)], \qquad B_{ij} = \exp[\alpha - \alpha_{ij}].
\eea
An alternative reference system is the Debye--H\"uckel model \cite{EbFoFi17}:
\be
V_{ij}(r)  =  Z_i Z_j k_{\rm B} T \frac{\ell}{r} + V'_{ij}( r) ; \quad V'_{ij}( r) = \infty \quad \text{if} \quad r < R_{ij}, \quad  \text{else} \quad V_{ij}'(r)  = 0, 
\ee
where $\ell$ is the Coulomb length and $R_{ij}$ is the contact distance. Nowadays, the Debye--H\"uckel model of charged hard spheres is a kind of standard model
with many succesful applications \cite{EFKS77,EFKS77_a,EbFeSa79,EbKr2023}.
On the other hand, in comparison with the Debye--H\"uckel model, the exponential potentials have several important advantages \cite{EbKr79}:
\begin{enumerate}[label=(\roman*)]
	\item The exponential potentials are rather smooth and possess a simple Fourier-transform. Therefore, it seems to be also well-suited for the treatment of all screening and collective effects.
	\item For systems with exponential potential, there exist strict Bogoliubov-type expansions  with respect to the plasma parameter.
	\item The existence of a simple screening equation permits cluster expansions around the 
	first exponential-type approximation similar to those for Debye--H\"uckel-type systems \cite{EbKr71}.
	\item As shown by Kelbg, the exponential potential is quite useful in applications to quantum plasmas~\cite{Kelbg62,Kelbg62a,Kelbg63,KelbgHoff64}.
\end{enumerate}

In general, in what follows we prefer to use exponential potentials with just one $\alpha$. We want to show that this basic case is already very flexible 
and permits, if adapted in the best possible way to the system, to reach excellent descriptions of real systems already in the first approximation.
The easiest way to map a binary system to an approriate $\alpha$-parameter is the following choice of the $\alpha$-parameter:
\begin{enumerate}[label=(\roman*)]
	\item $\alpha = \alpha_{+-}$. This choice is related to the so-called opposite charge approximation in plasma theory.
	\item  More appropriate may be the request that the exponential potential covers the first deviation of important thermodynamic and transport quantities from the limiting law, in an exact way. We show how we may go from a matrix system with many $\alpha$-parameters, to a reference system with just one parameter. The request in the case of thermodynamic functions leads to the choice
\bea
\frac{1}{\alpha} = \frac{\sum_i \sum_j e_i^2 e_j^2 (1/ \alpha_{ij})}{\sum_i \sum_j e_i^2 e_j^2}.
\label{pairav}
\eea
\end{enumerate}

A different, even more efficient way, which we introduce herein later, is averaging of the screening functions at the level of pairs similar to  equation~(\ref{pairav}). 
The most important advantage of the exponential potential as a reference system is based on the fact that the
Fourier transform is so simple:
\bea
{\tilde V}_{ij} (k)=  Z_i Z_j V(k) = Z_i Z_j  k_{\rm B} T \ell \Big[\frac{4 \piup}{k^2} - \frac{ 4 \piup }{k^2 + \alpha^2} \Big].
\eea
We note that our potential includes a weak short-range repulsion having a  Fourier-transform. In spite of the simplicity
of an extension by matrix terms, the screening problem is then much more difficult.
The question about the choice of a charge-independent $\alpha$-parameter is to decide from case to case that the leading aspect 
should be that the encounters of opposite charges are well described.

The question whether two potential models A and B are equvalent is not trivial. Most often one uses the rule 
that the two 2nd virial coefficients for the two potentials are similar or even identical. 
This, however, does not work for Coulomb systems, since the 2nd virial coefficients do not exist.
Therefore, we use the prescription that two potential models for the ionic interaction are equivalent, if the first deviations from
Debyes limiting law are equal.

We start as in \cite{EbKr79} from the case of no additional short-range corrections and use the Bogoliubov approximation
$F_{ij} (1,2) = 1 + g_{ij} (1,2)$:
\bea
g_{ij}(1,2) + \beta V_{ij}(1,2) + \beta \sum_k n_k \int V_{ik}(1,3) g_{jk} (2,3)\, \rd {\bf 3} = 0.
\eea
This approximation may be considered as the first term of a Bogoliubov-type expansion with respect to the plasma parameter \cite{Falkenhagen,Falkenhagen_a}.
In some sense, the Bogoliubov equation for the correlaton function $g_{ij}$ is related to the Ornstein-Zernike equation which is the basis of 
HNC and MSA-related theories \cite{Barthel}. We have to note, however, that the Bogliubov-type approach is historically 
the first systematic approach to Coulomb systems and is related too, but in detail it is quite different from MSA-type approximations 
\cite{Barthel}.

In the general case that $\alpha$ is a matrix, the correlation function was  obtained  by solving the integral equations \cite{Falkenhagen,Falkenhagen_a,EbFoFi17}. In the simplest case, that all $\alpha$ are equal, or that we use an averaged one, the solution is particularly easy and reads
\bea
g_{ij} = - Z_i Z_j \frac{\ell \alpha^2}{r (p^2 - s^2)} \big[\exp(-pr) - \exp(-sr)   \big] .
\eea
The parameters $s$ and $p$ are solutions of a 4th order polynomial. At very small densities, the solutions are
$s = \alpha$ and $p = \kappa = 1/r_D$. At intermediate densities we have \cite{Falkenhagen,Falkenhagen_a}
\bea
 p = \frac{\alpha}{2} \bigg[\sqrt{1 + 2 \frac{\kappa}{\alpha}} -\sqrt{1- 2 \frac{\kappa}{\alpha}}\bigg]; \quad s = \frac{\alpha}{2} \bigg[\sqrt{1 + 2 \frac{\kappa}{\alpha}} +\sqrt{1 - 2 \frac{\kappa}{\alpha}}\bigg].
\eea
We study now the thermodynamic functions for the case that we have only one $\alpha$. The Coulomb energy is 
\bea
U^C =  \frac{V}{2}k_{\rm B} T \sum_{ij} n_i n_j Z_i Z_j \int \frac{\ell }{r_{12}} g_{ij} (1,2)  \rd {\bf 2}=
 - k_{\rm B} T \sum_{ij} n_i n_j Z_i^2 Z_j^2 \ell^2  \frac{\alpha^2}{(p^2 - s^2)}\bigg[\frac{1}{p} - \frac{1}{s} \bigg].
\eea  
The most easy way to extend this formula to mixtures with parameter $\alpha_{ij}$ avoiding a matrix calculation is the averaging at the level of the screening functions of pairs instead of averaging the 
$\alpha_{ij}^{-1}$ as in equation~(\ref{pairav}) 
\bea
U^C =   - k_{\rm B} T \sum_{ij} n_i n_j Z_i^2 Z_j^2 \ell^2  \frac{\alpha_{ij}^2}{(p_{ij}^2 - s_{ij}^2)}\bigg[\frac{1}{p_{ij}} - \frac{1}{s_{ij}} \bigg].
\eea
Further results and applications will be given below. The success of the description by exponential potentials is based on the fact that an essential part of the short-range effects is included by an optimum choice of the exponential potential.
According to Yukhnovskii and Kelbg, we get for the contribution of the Kramers--Hellmann potential \cite{Yukhn02,Kelbg59,Kelbg59a} 
\begin{align}
&\beta F_{KY} = \beta F_{\rm id} - \frac{\kappa^3}{12 \piup} \tau\Big(\frac{\kappa}{\alpha}\Big) ; \qquad \beta p = n - \frac{\kappa^3}{24 \piup}\varphi\Big(\frac{\kappa}{\alpha}\Big),\\
&\varphi(x) = \frac{1}{x^3} \bigg(\frac{1}{4} X^3 - \frac{2}{3} X - \frac{3}{4 X}\bigg); \qquad 
\tau (x) = \frac{3 \alpha^3}{4 \piup \kappa^3}\bigg(Y^2 - 2 Y + \frac{2}{3} Y^3 - \frac{1}{2} Y^4 + \frac{5}{6} \bigg),\\
&x = \frac{\kappa}{\alpha}; \qquad X = \sqrt{1 + 2 \frac{\kappa}{\alpha}};  \qquad Y = 2 + X.  
\label{ringfct}
\end{align}
For the excess chemical potential of an ion and the mean activity coefficient there follows a simple form~\cite{GlaYuk52}
\bea
\mu_i^{\rm ex} = - \frac{e_i^2 \alpha}{4 \piup \epsilon_0 \epsilon_r k_{\rm B} T \kappa} \bigg[ 1 - \frac{1}{\sqrt{1 + 2 (\kappa/\alpha)}}  \bigg],\quad 
\log f_{\pm} = - \frac{|e_1 e_2|  \alpha}{4 \piup \epsilon_0 \epsilon_r k_{\rm B} T \kappa} \bigg( 1 - \frac{1}{X}  \bigg).
\eea
Remember that we have for the Debye--H\"uckel model for the mean activity and the free energy function
\bea
\ln f_{\pm}^{\rm DH} = - \frac{|Z_1 Z_2| \ell \kappa }{(1 + \kappa a) }; \qquad 
\tau_{\rm DH} (\kappa a) = 1 - \frac{3}{4} (\kappa a) + \frac{3}{5} (\kappa a)^2 -  \frac{3}{6} (\kappa a)^3 +  \dots 
\eea
For comparison of the Debye--H\"uckel reference system with the simple exponential system, we may  define $a$-parameter conjugated to our $\alpha$-parameter by $a = (3/ 2 \alpha)$ or in the case of mixtures by
$a_{ij} = (3/ 2 \alpha_{ij})$. 

Further corrections by a short-range and a hard core part of the potential may be included by the second virial contribution \cite{Barthel}.
This will be discussed below. The case of complex $\alpha$ which corresponds to oscillating potentials remains to be discussed. 

\subsection{Individual electrolytic conductivities}
As known already to Nernst, the effect of nonideality on the conductivity is structurally closely related to the effect on the osmotic pressure.
Including external electrical fields is an extra big chapter of the electrolyte theory \cite{Falkenhagen,Falkenhagen_a,Barthel} which we can only touch upon here, making use of the many parallels between individual osmotic coefficient and transport. As a matter of fact, not only historically but
also systematically, the extension of the equilibrium theory of electrolytes to weak external field effects is based on quite similar
tools of statistical physics. In our case, the relevant tools are the pair distribution functions and the tools of density and interaction. Here we haven't got enough  space to repeat
the transport theory \cite{Falkenhagen,Falkenhagen_a,EbMatrix24}, we give only the key ideas and present some extrapolations.

Under the influence of weak external electrical fields $E$, the pair distribution is split into an equilibrium and a non-equilibrium part which is a solution of differential equations formulated by Onsager. The perturbation
is different from zero only for oppositely charged ions and is known
in several approximations~\cite{EbFeSa79}:
in order to get the effects of nonideality on conductivity we have to calculate by standard methods~\cite{Falkenhagen,Falkenhagen_a}
the relaxation force and the electrophoretic force and the corresponding contributions to
the specific conductivity~\cite{EbFeKr21,EbFeKr21a}:
\bea
\frac{\sigma_i}{\sigma_{i0}} = {\bar \alpha}_i + S_i^{\rm elp} + S_i^{\rm rel}; \qquad \sigma_{i0} = \frac{n_i e_i^2}{\rho_i},
\eea
where $\sigma_{i0}$ is the ideal conductivity and $\rho_i$ is the friction parameters of the ions.
The first term reflects the decrease of conductivity due to association. This is the Arrhenius effect which is parallel to the effect of the osmotic coefficient. As far as the association is weak, we may just take over the earlier found expression. The next correction which is dominant is due to hydrodynamic interactions and is called electrophoretic effect.
For the estimation of the electrophoretic force action on the ion $i$ may use the expression for the Coulombic energy per ion 
since  ($\eta_0$ --- viscosity) \cite{Falkenhagen,Falkenhagen_a}:
\bea
S_i^{\rm elp} =   - k_{\rm B} T \frac{\rho_i  Z_i^2}{3 \piup \eta_0}  \sum_{j} n_j Z_j^2 \ell^2 \bigg[\frac{1}{p} - \frac{1}{s} \bigg].
\eea
Note that in the case that the the short-range components of the forces are species-dependent, the solution for the correlation function is more complicated and strictly speaking needs a matrix theory.

Our approach will be applied now to conductivities using, in particular, the fact that individual conductivities are structurally related to the individual energies and osmotic pressures.
We show that we can proceed this way to concentrations of interest for the description of measured conductance data. Further we also  include some applications of the exponential potential to quantum statistics which go back to Kelbg \cite{Kelbg62,Kelbg62a}.
We use as measures of concentration the molar concentration $c_i$ in mol/dm$^3$ in the solute as well as the density $n_i = N_i/V$ of the species $i$.
The first term reflects the Arrhenius effect, i.e., a decrease of conductivity due to association. As far as field effects are
weak, we may just take over the expression for the degree of ionization given above. The next correction is due to hydrodynamic interactions. Its contribution to the specific conductivity is negative and can be expressed by the electrical part of the
electrical energy density.
This way, two nonideality contributions are already known from equilibrium results.
Expressions for the relaxation part of conductivity are known for ions with non-equal contact distances $a_{ij} = R_{ij}$:
\bea
S_{i}^{\rm rel} = - \frac{\ell \kappa q}{3 (1-q) }\sum_j \zeta_j^0 \bigg(1 +\frac{|Z_i|}{|Z_j|}\bigg)\bigg(\frac{1}{1 + \kappa a_{ij}} -  \frac{\sqrt{q} }{1 + \sqrt{q} \kappa a_{ij}}\bigg) .
\eea
The free parameters are here the contact distances and the friction constants $\rho_i$, which are related to the Stokes radii and to
the limiting molar conductivities $\sigma_{i0}$ by classical relations and $q = 1/2$ \cite{Falkenhagen,Falkenhagen_a}. We approximated here the relaxation part of the conductivity as a sum of binary contributions and refer to this as a partial fraction representation.

\begin{figure}[htbp]
\begin{center}
\includegraphics[scale=0.7]{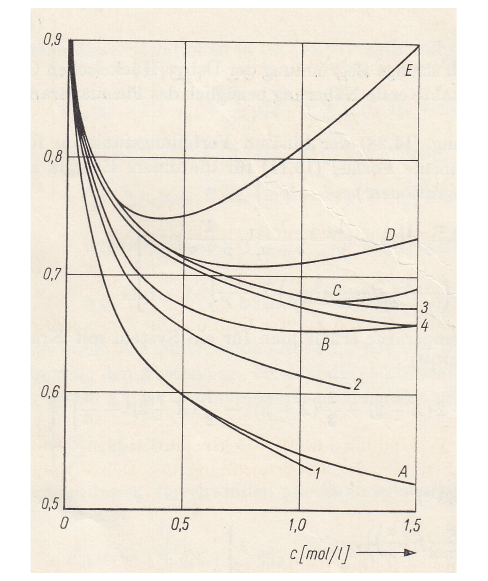}
\includegraphics[scale=0.85]{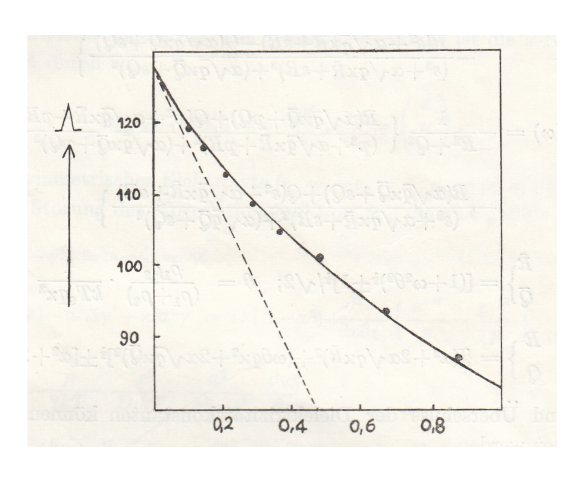}
\caption{Left-hand: mean activity coeffiient according to Kelng \cite{Falkenhagen,Falkenhagen_a} for different parameters $\alpha^{-1} = 1.85, 2.50, 4.09, 4.50$~\AA~(from below)  as well as several exprimental curves for LiOH--A, NaOH--B, NaBr--C HCl--D  \cite{Falkenhagen,Falkenhagen_a}. Right-hand: the conductivity of NaCl with the Kelbg's theory for this  potential with adapted $\alpha^{-1} = 3.1$ \AA~\cite{Falkenhagen,Falkenhagen_a}.}
\label{KelbgNaCl}
\end{center}
\end{figure}

The corresponding expressions for the exponential potential are also known in part. Particularly easy is the electrophoretic part which is expressed by the  known Coulomb $u_i^{\rm el}$  given above. 
In order to find the relaxation part of conductivity for more general exponential potentials, we have to calculate the perturbation of the correlation function in the external field (here in $x$-direction). For our 
exponential potential   we find
\bea
g_{12} (r) \sim \frac{\partial}{\partial x} \bigg\{ A_0 \bigg[ \frac{\exp(-p r)}{r} -  \frac{\exp(-s r)}{r} \bigg]
+ A_1  \bigg[\frac{\exp(-p' r)}{r} -  \frac{\exp(-s' r)}{r}    \bigg]\bigg\}.
\eea
The calculation of the relaxation  term is easy for $B_{12} =1$; the two relevant roots are \cite{Falkenhagen,Falkenhagen_a,Kelbg59,Kelbg59a}:
\bea
p' = \frac{\alpha}{2} \bigg(\sqrt{1 + 2 \sqrt{q} \frac{\kappa}{\alpha}} -\sqrt{1 - 2 \sqrt{q} \frac{\kappa}{\alpha}}\bigg); \quad s' =  \frac{\alpha}{2} \bigg(\sqrt{1 + 2 \sqrt{q} \frac{\kappa}{\alpha}} +\sqrt{1 - 2 \sqrt{q} \frac{\kappa}{\alpha}}\bigg). \nonumber
\eea
Based on this result, we may obtain the conductivity for the exponential potential.
An application to NaCl in aqueous solution is shown in figure~\ref{KelbgNaCl}  representing the results due to Kelbg. We see a very good agreement of theory with the data up to $1$~mol/l~\cite{Falkenhagen,Falkenhagen_a}.  Kelbg found as adapted parameter $\alpha^{-1} = 370$~pm. 
One may be sure that the model with more than 1 adaptable $B_{ij}$-parameters describes for NaCl the data even up to $1.5$~mol/l. In figure~\ref{KelbgNaCl} we show the theory for charged spheres, corresponding to the parameters $R_{\rm NaNa} = 479$~pm, $R_{\rm NaCl} = 350$~pm, $R_{\rm ClCl} = 360$~pm found by fitting contact distances in \cite{EbFeKr21,EbFeKr21a}.  
The model of charged hard cores is in reasonable agreement with the data only up to $c = 0.2$~mol/l. Beyond this concentration, the theoretical curve starts to deviate from the data for NaCl at $c > 0.3$~mol/l, although the exponential potentail model still fits well. We may conclude that the exponential potential describes the effective forces somewhat better.  
We note that the Pauling radii of the corresponding ions are $R_{\rm Na} = 95$~pm, $R_{\rm Cl} = 167$~pm, which means that the fitted contact distances substantially deviate from the sums of the Pauling radii. In conclusion, we may say that the exponential potential 
used by Kelbg and Yukhnovskii is well suited, and probably a combination of exponential potential and hard core potentials has 
best perspectives.

Our figure~\ref{KelbgNaCl}  which is due to Kelbg \cite{Falkenhagen,Falkenhagen_a}, demonstrates an excellent description of the data up to $1$~mol/l for the mean activity coefficients and for the conductivity. Anyhow, we see that the given theory for the exponential potential given above provides a most  simple and useful first approximation
which agrees well with more advanced theories of the ionic conductivities discussed above \cite{EbFeKr21,EbFeKr21a}. 
A different successful strategy is to combine an exponential potential with one free parameter with a hard core potential.
In all cases it is highly recommended to use for the fit not only one set of data for a specific property
but much better several data series for different properties, e.g., activities and conductivities.
Note that conductivities permit very precise measurements and are more sensitive to association and to anion-cation interactions.

\subsection{Virial expansions for classical charged particles}
More recently several thermodynamic and transport properties have been calculated for models of charged partcles. 
For simplicity, here we use charged spheres  with adaptive
non-additive contact distances~$R_{ij}$~\cite{EbFeKr21,EbFeKr21a}. 
The long-range part $V_{ij}$ is identified with the exponential potential of Kramers and Hellmann and consider the contact distance $R_{ij}$ as an adaptable parameter. This is the route we considered
in the papers  \cite{EbFeKr21,EbFeKr21a,EbKr2023}.
The statistical theory applied to our model potentials is based on the classical work of Mayer, extended by Friedman
 \cite{Friedman} and, in particular, by the collaboration Yukhnovskii--Kelbg~\cite{Yukhn80,EbKe66,EbKe66a}. 
This theory based on cluster expansions provided several exact results for the thermodynamic functions valid for small ionic densities.
General expressions from statistical thermodynamics for the cluster contributions $S$ to the negative
free excess energy of hard charged spheres read \cite{Falkenhagen,Falkenhagen_a,Friedman,Yukhn80,EbKe66,EbKe66a}
\begin{align}
F_{\rm ex} = F_{DH} + F_2 + F_3 + \ldots = - k_{\rm B} T V \bigg[\frac{\kappa^3}{12 \piup} \tau\bigg(\frac{\kappa}{\alpha} \bigg)  + \sum_{i,k} S_{i}^{(k)} \bigg],
\end{align}
where $\tau(x)$ is the so-called ring function. The cluster integrals are expressed by the cluster pair functions~$\psi$ which are in 
Kelbg--Yukhnovskii approximation 
\begin{align}
\psi_{ij} = \exp[g_{ij} - \beta V'_{ij}] -1 - g_{ij}; \qquad  \tau(x) = 1 - \frac{9}{8}x + \frac{3}{2} x^2 - \frac{21}{12} x^3 + \ldots
\end{align}
The sums should be extended over the species of ions $i$ and all orders of clusters $k$.
The strong coupling contributions of ions $i$ read in the cluster order $k = 2,3$ \cite{Falkenhagen,Falkenhagen_a,Friedman,Yukhn80,EbKe66,EbKe66a}
\begin{align}
S_i^{(2)} &=  \frac{1}{2} n_i \sum_j n_j \int \rd {\bf r}_j \textbf{g}[\psi_{ij} - \frac{1}{2} g_{ij}^2\bigg], \nonumber\\
S_i^{(3)} =  \frac{1}{2 \cdot 3} n_i \sum_{jk} n_j n_k &\int \rd {\bf r}_j\, \rd {\bf r}_k \big[\psi_{ij} \psi_{ik} \psi_{jk} + g_{ij} \psi_{ik}\psi_{jk} + g_{ik} \psi_{jk}\psi_{ij} + g _{jk} \psi_{ij}\psi_{ik}\big]. 
\end{align}
The basic elements is the cluster function $\psi_{ij}$ which is of the order $O(e^4)$ in the interactions.
We see here that the 2nd cluster integral starts with the order $e^4$ and the 3rd virial coefficient with the order $e^8$ \cite{EbKr2023}.

Several  values for the contact distance were fitted to the data \cite{EbFeKr21,EbFeKr21a,EbKr2023}.
We notice important deviations from additivity of individual radii. 

\begin{figure}[htbp]
\begin{center}
\includegraphics[scale=0.6]{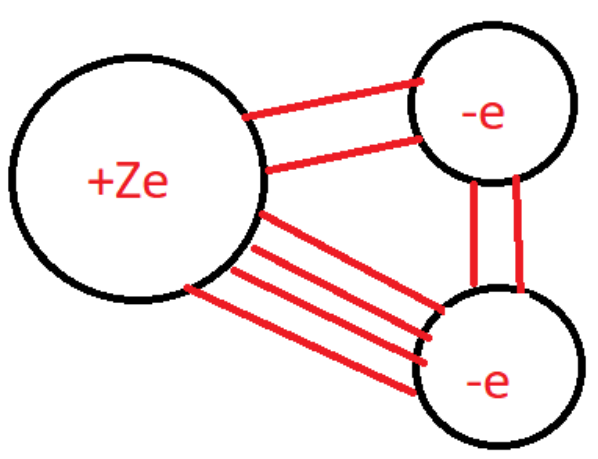}
\includegraphics[scale=0.6]{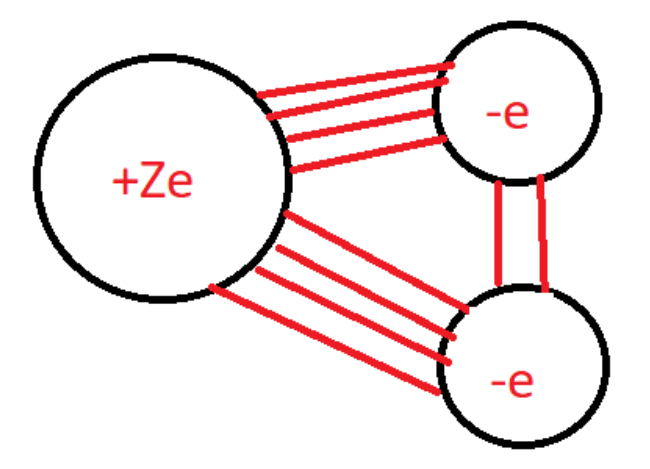}\\
\caption{(Colour online) Typical 3-particle diagrams which we study here. The diagrams show one 2-fold charged positive ions 
as Mg$^{++}$ or He$^{++} $ and 2 negative single-charged ions as Cl$^-$ or electrons connected by interaction bonds.
The 2-fold and 4-fold lines stand for 2nd order and 4th orders of the expansions of the $\psi$-functions.}
\label{Mayer}. 
\end{center}
\end{figure}

{\bf Calculation of triple integrals and mass action functions:}
We come now to the derivation of mass action constants for the triple formation. Our basic assumption is that the convergent part of the cluster integrals is directly related to the mass action 
function of triple formation. More presicely, analyzing the triple cluster integral we have to select those parts which are non-negative, grow exponentially for $\epsilon k_{\rm B} T \rightarrow \infty$ 
and depend only on the energy units, which are classically the Bjerrum energy $U_{Bj} = e^2/ 4 \piup \epsilon \epsilon_0 R_\pm$ or the Rydberg energy Ry in the quantum case.
In figure~\ref{Mayer} we show several Mayer-like diagrams which are possible candidates for bound state contributions of 3 charged particles of type
$(2+,-,-)$  since they contain at least one 4-fold $(+-)$-bound. From the 2-particle bound state problem, we know that a 4-fold $(+-)$-bound provides a tight $(+-)$-binding of two particles. The investigations of
the diagrams with 3 or 4 nodes with only one 4-fold binding shows, however, that the integrals are, according to Friedmans estimates, evidently divergent. Including, however, two 4-fold $(+-)$-bounds, we arrive at the  graphs (the lower diagrams) which are for sure convergent even without
screening. This is the reason why we may include them in the classical and in the quantum case
as the lowest order binding contributions of 3 or more particles.
The basic assumption of the present paper is that the mass action constant is related to an asymptotic part of the 3rd virial coefficient, which 
should be convergent and non-negative and a
exponentially growing function of the reciprocal Bjerrum temperature ${\bar \beta} = U_{Bj} / k_{\rm B} T$ or Rydberg temperature 
${\tilde \beta} = Ry / k_{\rm B} T$. The technique of concrete evaluation of the virial coefficients is
quite complicated \cite{Barthel,Friedman}. The method we used in \cite{EbFeKr21,EbFeKr21a,EbKr2023} is based on expansions with respect to $e^2$. We  provided 
in \cite{EbKr2023} only the coefficient of $e^{10} \sim {\tilde \beta}^5$. Here we generalize the approach and also obtain the coefficients of 
$e^{8} \sim {\tilde \beta}^4$.  In the fugacity series which is relevant for the determination of the mass action coefficients there appear reducible and irreducible terms (see  \cite{BaEb71} and figure~\ref{Egraphs0}), which both should be evaluated.

We may give this way the missing order $ {\bar \beta}^4$ which is quite relevant. 
In the integrals there appear a strong coupling exponential function $\psi_{ij}$ analyzed in detail in \cite{EbKr2023} which is of a higher order $O(e^4)$ 
in the interactions.
 One of the reducible terms in the 3rd virial coefficient is just a square of the second virial coefficient.
This provides a contribution $\delta K^{(3)}(T) \sim [ S_i^{(2)}]^2$. With a correct normalization we get 
$\delta K^{(3)}(T) \sim 50 {\bar \beta}^4$.
For calculation of the irreducible graphs we have to solve 2-center integrals of the following type
\begin{align}
\delta K^{(3)}(T) \sim  \int_1^{\infty} \rd r'\, \left(1/r'\right)^{n-1} \int_1^\infty \rd s' \, (1/s')^{m-1} \int_{|r'-t'|}^{r'+s'} \rd t' \, \left(1 / t'\right).
\end{align}
The case $m=n=4$ was calculated in  \cite{EbKr2023} leading to $a_5 = 3.1$. We add here the calculation for the case $n = 2$, $m=4$
finding the integral 
\begin{align}
C_{n=2, m=4} = \piup^2 \int_1^{\infty} \rd r'\, \left(1/r'\right)^3 \int_1^\infty \rd s' \, (1/s')^2 \  \simeq  \frac{\piup^2}{2} \simeq 4.93.
\end{align}
This provides the coefficients which are the sum of reducible and irreducible graph (48.92)  
\begin{align}
a_4 = (35.1 + 48.9) = 84     ; \qquad a_5 = 15.3.
\end{align}
Completing the previous paper we calculated the lowest order in the Taylor expansion of the triple mass action constant which permits 
to find a complete expression for the triple mass action constant 
\begin{align}
K^{(3)}  = a_4 {\bar \beta}^8 + a_5 {\bar \beta}^{10} + \sum_{k=6} \frac{(7 {\bar \beta})^{2k}}{2 (2 k)!}
\end{align}
for all (reduced) temperatures. 
Note that in comparison with reference~\cite{EbKr2023}, the new values for the triple association constant are larger than the 
results obtained earlier for MgCl$_2$ and for Na$_2$SO$_4$.
Correspondingly, the new estimates give also larger values for the degree of triple association using the expressions
\begin{align}
\delta^{(3)}  = 2 n_+ n_{-}^2  f_{\pm}^2 K^{(3)} (T); \qquad \ln f_{\pm} =  - \frac{z_+ z_-}{2 }\frac{\kappa \ell}{(1 + \gamma_{\pm})}.
\label{alpha0}
\end{align}
Here, $f_{\pm}$ are the activity coefficients in electrical oppositely charge approximation \cite{EbFeKr21,EbFeKr21a}.

\section{Quantum statistics and cluster expansions}

\subsection{Kelbg potential and cluster expansions}
At the end of the 60th and in particular after the first visit of Ihor Yuknovskii to Rostock University in 1968, quantum-statististical investigations
were included into the collaboration  beween the two schools. The collaboration led an exchange of students and aspirants and to many remarkable results \cite{Albehrendt,GoKr89,EbHoKe67,EbHoKe67a,EbHoKe67b,KelbgEb70,KelbgEb70a,Kelbg72,EKK76}. 
One of the tasks was to extend the virial expansions in the canonical ensemble developed by Ihor Yukhnovskii \cite{Yukhn58,Yukhn80} and G\"unter Kelbg \cite{Kelbg62,Kelbg62a}.
In particular, the development of the method of fugacity virial expansions which we will present now, started in 1969 at 
a seminar in the Institute on the Dragomanov street in Lviv founded by Ihor R. Yukhnovskii. Since Prof. Yukhnovskii was of a high opinion about the prospects of this new ``ansatz'' based on fugacity virial expansions, he proposed to publish the 2 lectures in Lviv as Preprints of the Ukrainian Institute of Theoretical Physics \cite{KelbgEb70,KelbgEb70a}.
Here we summarize how the method presented for the first time in 1969 in a seminar at the Institute in Lviv, has developed over the years.
In order to take into account the screening effects into statistical expansions we need to sum up an infinite number of ring-type diagrams
in the grand-canonical ensemble.

For quantum-statistical Coulomb systems, the pair potential consists, the same as for the classical case, of a long-range part $V_{ab}$ which contains a Coulomb
tail and has a Fourier transformable part and a short-range part. Following the work of Kelbg \cite{Kelbg62,Kelbg62a,BoEbal22}, the quantum form of the exponential potential is  in the first approximation given by
\bea
 V_{ij}^K(r) = \frac{e_i e_j}{4 \piup \epsilon_0 r} \left[1 - \exp \big(- \alpha^K_{ij} r \big)\right]; \quad  a_{ij}^K = \frac{2}{3 \alpha_{ij}^K} =  \frac{\sqrt{\piup}}{9} \lambda_{ij}; \qquad \lambda_{ij} = \frac{{\bar h}}{2 m_{ij} k_{\rm B} T}.
\eea
Here, we introduced, besides the $\alpha$-parameter, a corresponding distance parameter $a_{ij}^K$; which is a kind of effective hard core diameter for charges in a quantum plasma \cite{EbHoKe67,EbHoKe67a,EbHoKe67b,EKK76,EbFoFi17}.
Note that in many applications instead of a manifold of $\alpha$ parameters, only one $\alpha_{\pm}$ is used. This is the simple opposite-charge approximation (OPA), which works well for two component plasmas in the region of partial ionization \cite{EbFoFi17}. The quantum potential has a finite height in $r = 0$.
A more precise quantum-statistical caculation leads Kelbg to the expression \cite{BoEbal22}
\begin{equation}
 V_{ij}^K(r) = \frac{e_i e_j}{4 \piup \epsilon_0 r} \left\{1 - \exp \left(\frac{r^2}{\lambda_{ij}^2} \right)
+ \sqrt{\piup} \frac{r}{\lambda_{ij}}\left[1-\Phi\left(\frac{r}{\lambda_{ij}}\right) \right] \right\}.
\end{equation}
We used here the standard definition of the error function $\Phi(x)$. 
For the Fourier transform, we get~\cite{Kelbg62,Kelbg62a,KelbgEb70,KelbgEb70a}
\bea
V_{ab}^K (t) =  \frac{e_a e_b}{\epsilon_0 t^2} M\bigg(1, \frac{3}{2}, -\frac{1}{4}\lambda_{ab}^2 t^2 \bigg),
\label{Ftransf}
\eea
where $M = {_1}F_1$ is  the well known confluent hypergeometric function, which is also called Kummer function. This first result by Kelbg which still neglects the exchange effects had been given 
already in 1963. The  Kummer function may be represented in integral form or using a Taylor expansion  or alternatively the asymptotics shows a decay like $x^{-2}$:
\bea
M\bigg(1, \frac{3}{2},-x^2\bigg) = \frac{1}{2} \int_0^1 \rd u \, \frac{\exp(- u x^2)}{\sqrt{1 - u}}  = 1 - \frac{1}{6}x^2 + \frac{1}{30} x^4 + \ldots \simeq \frac{1}{(1 +  x^2/6)}.
\eea
We see that the Kummer function may be approximated by a simple rational function corresponding to equation~(\ref{eq_1}), i.e., to an exponential form of the potential.
This way  the rational approximation leads us back to the previous exponential approximation with 
$$
\alpha_{ij}^{-1} = (\sqrt{\piup}/6) \lambda_{ij}.
$$
This indeed shows a close relation between 
the full Kelbg quantum potential and the exponential potential, inspite of the quite different mathematical form. We have shown that the Fourier transforms of both potentials are quite similar and differ only for large Fourier vectors. Therefore, in applications, which are based on the Fourier transform as the screening procedure, we may replace the original Kelbg potential by the exponential potential and use the Kelbg--Yukhnovskii screening theory.

Besides we note that Kelbg also included the exchange effects and provided the Fourier transforms 
using hypergeometric functions, although the 
back transform to real space has not yet been found. A generalization to nondiagonal representations is the so-called Filinov approximation which is of much interest for numerical
applications in path integral calculations \cite{EbFoFi17,Fortov20}.

\begin{figure}[htbp]
\begin{center}
\includegraphics[scale=0.4]{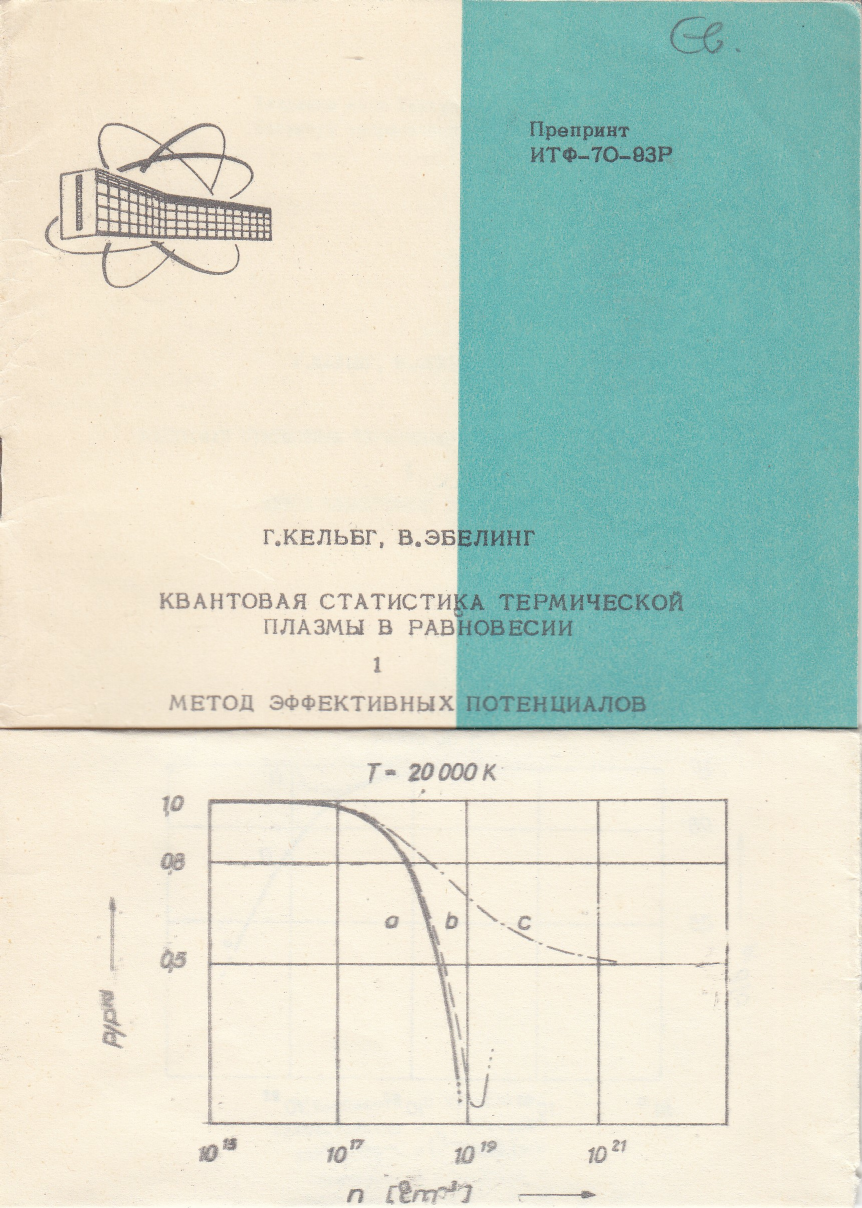}
\includegraphics[scale=0.335]{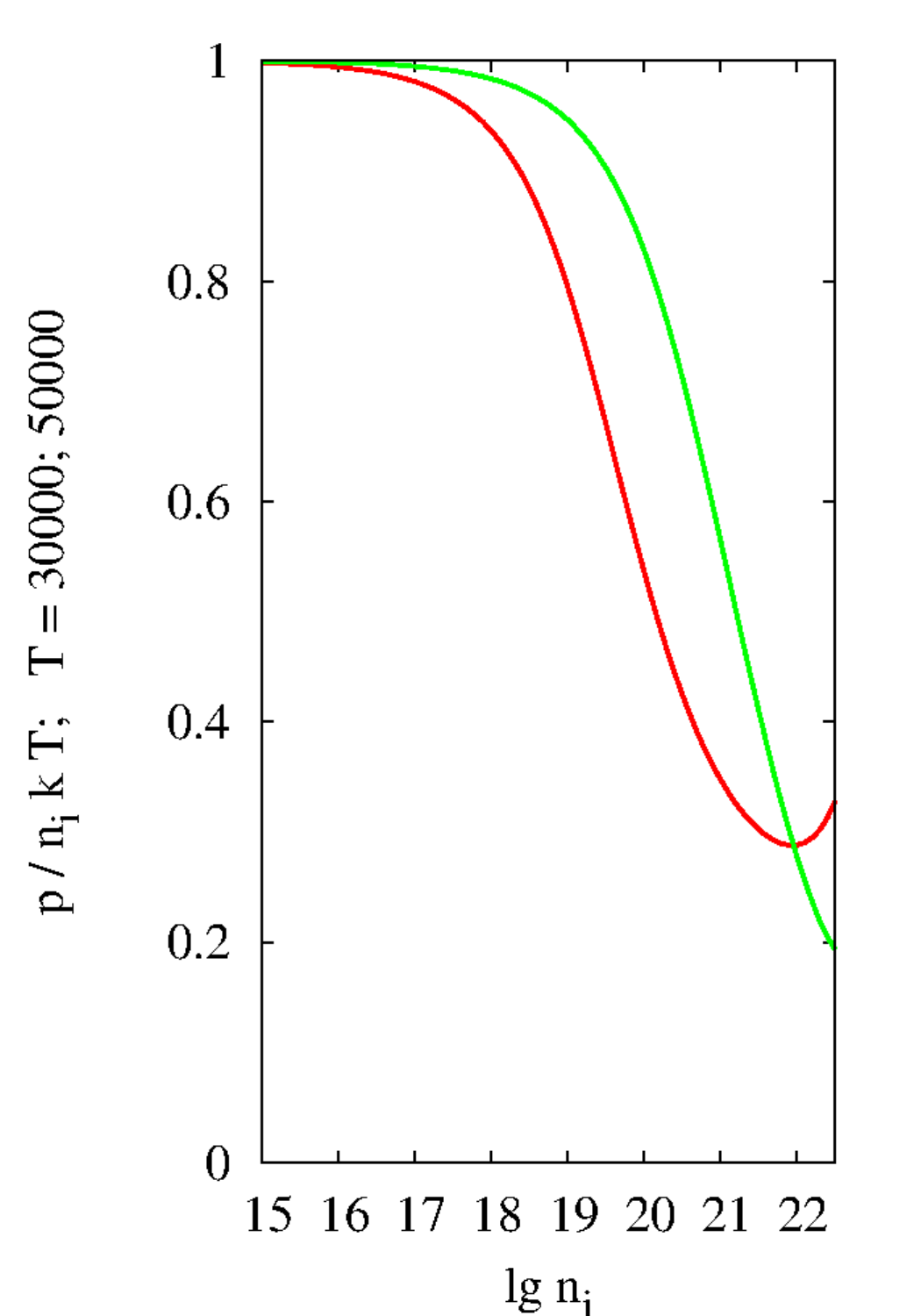}
\end{center}
 \caption{(Colour online) Left-hand: equation of state (EoS) of hydrogen plasmas from a fugacity expansion (upper thin curve c) in comparison with the density expansion (lower fat curve a) and the cover of the Preprint ITF-7093P (Kiev, 1970)~\cite{KelbgEb70}. Right-hand: the EoS of hydrogen plasmas calculated by using Pad\'e approximations \cite{EKK76,KKER86} for $T = 30\,000$ K (red curve) and for $50\,000$ K (green curve) as a function of ${\rm log} (n_i)$.
}
\label{qupress11}
\end{figure}
The fact that the main part of the potential  $V^K_{ab}({\bf r})$ has a Fourier transform
\bea
V^K_{ij}({\bf r}) = \frac{1}{V} \sum_k \exp(-\ri {\bf k}{\bf  r})\, V^K_{ij}({\bf k})
\eea
suggests a special mathematical technique
for the evaluation of the classical or quantum partition function which is based on the method of collective variables.
This method was developed for Coulomb systems suggested first by Bohm, Pines and Zubarev 
and worked out then by Yukhnovskii and Kelbg. This method works in the classical as well as in the
quantum case and typically leads to the following structure of the general partition function \cite{Yukhn58,Kelbg62,Kelbg62a}:
\bea
\log Z(T,V,N) = \log[Z_{\rm id} Z_{0}]  + \log Z_{\rm ring}(T,V,N)  + \sum_{n=2}^{\infty} \log Z_n (T,V,N).
\label{series}
\eea
Here, $Z_0$ is the lowest order term which is of the order $e^2$ in the interaction and is
connected with the contribution of the zero wave vector ${\bf k} =0$ to the Fourier transform
\cite{Yukhn80,Brilliantov}.
This contribution is denoted as the self-energy of the long-range field \cite{Brilliantov}.
It comprises the Hartree--Fock terms as well as possibly some
ideal quantum terms (in dependence on the definition of the term $Z_{\rm id}$).
The following ring term comprises a set of contributions
typical of Coulomb systems which describe the screening effects and contain the Debye contribution
and generalizations. The contributions $Z_0 (T,V,N)$ and $Z_{\rm ring} (T,V,N)$ are mainly determined by the Fourier transform
of the long-range part of the effective potential and are essentially determined by contributions of order the 
$e^4$ in the interaction. This procedure traditonally performed in the canoncical ensemble.
In connection with the task to treat the bound states in an effective way, in lively discussions in the 
Yukhnovslii seminars in 1969 in his Institute in the Uliza Dragomana the idea arose  to transfer
the whole procedure
to the grand canonical ensemble \cite{KelbgEb70,KelbgEb70a}.

This way this fruitful new method was first demonstrated in the mentioned seminars in Lviv under the auspices of 
Ihor Yukhnovskii \cite{KelbgEb70,KelbgEb70a,BaEb71}. Extensive applications to real plasmas were given later by Rogers and DeWitt in the Livermore group  \cite{Rogers,Rogers_a,Rogers79,Rogers79_b,Rogers01}. We explain here only the basic ideas and several analytical results. 
In order to have a closer relation between fugacities and densities, from now on we use the fugacities which are not dimenisonless as above but have a dimension of densities. For the definitions
of new fugacities  $z_i$ we use the relation with the chemical potentials $\mu_i$ by
\be
z_i = \frac{(2 s_i + 1)}{\Lambda_i^3} \exp(\beta \mu_i); \qquad \Lambda_i = \frac{h}{\sqrt{2 \piup m_i k_{\rm B} T}}.
\ee
In terms of these new fugacities, we may derive a statistical expansion of the pressure \cite{KelbgEb70,KelbgEb70a,BaEb71,EKK76,KKER86}
using as in \cite{BaEb71} the Debye screening.  However, much better is an approximation by  screening  the  full exponential potential as in the classical case.
For the ring contributions to the partition function, according to \cite{Kelbg62,Kelbg62a} for one-component plasmas we find
\bea
 \log Z_{\rm ring}(T,V,N) = \frac{V}{4 \piup^2} \int_0^{\infty} \rd t\, t^2 \{w(t) - \ln [1 + w(t)]\}; \quad w(t) =\frac{\kappa^2}{4 \piup} {\tilde V}^K ({\bf t}).
\eea
The ring expression reduces in zeroth approximation to the Debye result. Here, we include the higher orders
corresponding to the full exponential potential which leads to an substantial improvement. In a first approximation, we use the exponential approximation for the Kelbg potential and identify all elements of the matrix $\alpha_{ij}$ by just the one cooresponding to the pair of opposite charges
$\alpha = \alpha_{ij} \rightarrow \alpha_{\pm}$. In this approximation, we avoid the matrix theory. 
A charging procedure in connection with the graph expansion leads to 
\bea
p_{\rm ring} = - \frac{K^3}{24 \piup} {\tilde f} \big(\frac{K}{\alpha}\big); \qquad {\tilde f} = 1 - \frac{3 \sqrt{\piup}}{16} (K \lambda_{\pm}) + \frac{1}{10} (K \lambda_{\pm})^2+ \ldots 
\eea
The ring function has been expressed here by an expansion \cite{EbKrRo12,EbKrRo12_a}. 
The full expression for the pressure consists, apart from
the Fermi--Dirac part, of the ring term and the cluster series.
\bea
\beta p = \beta p_{\rm FD} + \beta p_{\rm HF}  + \beta p_{\rm ring} 
 + \sum_{a,b} z_a z_b b_{ab} (K) + \sum_{a,b,c} z_a z_b z_c b_{abc}(K) + \ldots
\eea
As the basic statistical quantity we consider the grand potential $\Omega = F - \sum N_i \mu_i$ which neglecting the fluctuations corresponds to 
the pressure $\Omega = - pV$. The grand potential $\Omega $ is a thermodynamic potential if the variables are $T,V$ and the chemical potential $\mu$ or the fugacity $z=\exp(\beta \mu)$.
Note that the sums run here over all species. The ring function is of the same form as in the classical case equation~(\ref{ringfct}).
However, we have a grand-canonical screening quantity, instead of the canonical
\bea
K^2 = \sum_ b K_b^2, \quad  K_b^2 =4 \piup \ell Z_b^2 z_b. 
\eea
Note that we may have a more complicated dependence on the masses. As already mentioned, we simplified here and have only a 
dependence on the relative mass $m_{\pm}$ of oppositeley charged pairs~(OPA)~\cite{EbFoFi17,EbKrRo12,EbKrRo12_a}.

Completing the earlier studies we are mostly interested in the bound state contributions 
which are contained in the virial coefficient. This was already demonstrated  by Kelbg and this author in in the mentioned lecture in Lviv 1970 and is seen in figure~\ref{qupress11}. As  shown in these figures, the relative pressure
has characteristic shoulders which are signals of changes of the effective particles numbers in the system. 
In order to study these effects in detail, we first study a simple 
model system created by cutting all Coulomb forces at a finite distance $r_{\rm cut}$,
assuming $V_{ij} (r)= 0 $ if  $r > r_{\rm cut}$. This way we avoid the divergences connected with the long range Coulomb interaction, which are regularized considering the screening effects but still keep the essence of the bound states. We introduce new fugacities ${\tilde z}$ which are normalized  to converge at low densities to the densities at finite dilution \cite{EKK76}.
As is well known, cluster expansions wihtin the grand ensemble lead to the  series in terms of fugacities, which have a similar shape
as the series in terms of densities but contain more terms stemming from reducible diagrams. We want to show here that
expansions with respect to fugacities may be more appropriate for the description of the bound states than the well known virial expansions with respect to densities. We use here fugacities $\tilde z$ which are normalized in the way that they converge to densities at infinite dilution.
\begin{eqnarray}
 \beta p 
 &=&  \sum_{j=1}^\infty b_j \tilde{z}^j = \frac{1}{V} \left( \tilde{Q}_1 \tilde{z} + \frac{1}{2!} \tilde{Q}_2 \tilde{z}^2 + \dots \right) \nonumber \\
 &-&\frac{1}{2V} \left( \tilde{Q}_1 \tilde{z} + \frac{1}{2!} {\tilde Q}_2 {\tilde z}^2 \dots \right)^2
 + \frac{1}{3V} \left(\tilde{Q}_1 \tilde{z} + \frac{1}{2!} {\tilde Q}_2 {\tilde z}^2 \dots \right)^3.
\end{eqnarray}
We see that the coefficients of the order $ \tilde{z} ^N$ are  essentially given by $N$-particle traces.
By comparison of equal powers ${\tilde z}^k$, we get
\begin{eqnarray}
 b_1 = \frac{1}{V} \tilde{Q}_1 , \quad  b_2 = \frac{1}{2 V} \left(\tilde{Q}_2 -\tilde{Q}_1^2\right),  \quad 
 b_3 = \frac{1}{6 V} \left(\tilde{Q}_3 - 3 \tilde{Q}_1 \tilde{Q}_2 + 2\tilde{Q}_1^3 \right),\, \ldots\,.
\end{eqnarray}

\begin{figure}[htbp]
\begin{center}
\includegraphics[scale=0.5]{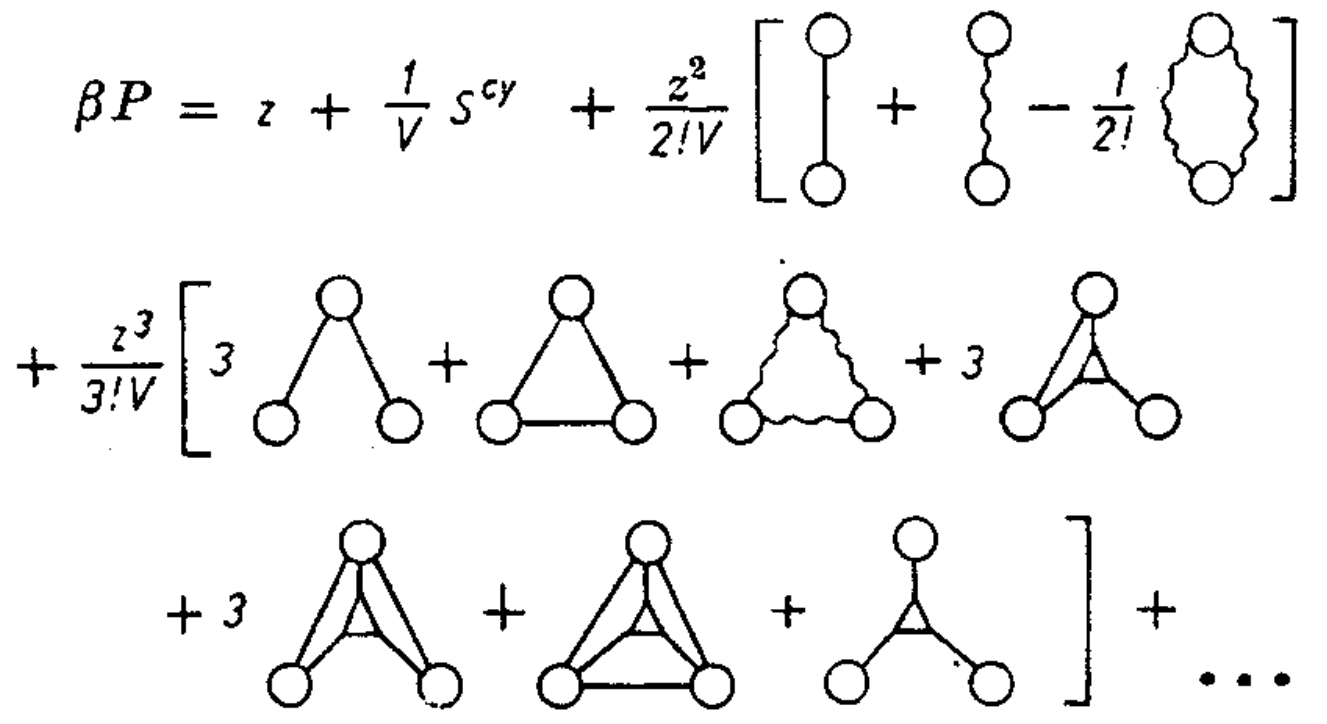}
\caption{Graphs representating the fugacity expansion of the pressure in a Mayer-like Kelbg--Morita 
representation. The coefficient of $z^3$ corresponds to reducible and irreducible graphs which represent the 3rd virial coefficient. 
The convergent part of these diagrams provides the partition function of helium. The wavy lines represent Coulomb-
(or Kelbg-) interactions and 
the normal lines stand for $\psi$-functions like in equations (23--24)   \cite{BaEb71}. 
}
\label{Egraphs0}
\end{center}
\end{figure}

The general relations follow from Thiele's semi-invariants. The partition functions which are constituents of the virial coefficients are expressed by
the  $s$-particle traces
\bea
{\tilde Q}_s (T,V) = {\rm Tr}_{1,2,..s} [\exp (-\beta H_s)] = {\rm Tr}_{1,2,..,s} [\exp( - \beta {\bar h}^2 \tau_s - \beta e^2 \varphi_s)],\\
H_s = T_s + V_S, \qquad \tau_s =  T_s / {\bar h}^2,      \qquad          \phi_s = V_{s} / e^2 .
\label{eq_44}
\eea
In this way all our partition functions ($Q$-functions) have the structure of the basic trace function equations~(\ref{eq_46}--\ref{eq_48}) and may be represented 
graphically as in figure~\ref{Egraphs0}. We remember the relation
\bea
 (-1) \beta e^2 \varphi_s = \sum_{1<i<j<s} \frac{\xi_{ij}}{|{\bf r}'_{ij}| };  \qquad  {\bf r'}_{ij}
= \frac{{\bf (r_i - r_j)}}{\lambda_{ij}},  ; \qquad \xi_{ij} = - Z_i Z_j \frac{\ell}{\lambda_{ij}},
\label{eq_45}
\eea
and see that the  contributions to equations~(\ref{eq_44}--\ref{eq_45}) after resolvent expansion have the following structure  
\bea
\tilde{Q}_s(T,V) =  \sum_{1 \leqslant i \leqslant j \leqslant s} \sum_n (-1)^n \xi_{ij}^n  \frac{1}{2 \piup \ri} \int_c \rd z \exp(- \beta z) {\rm Tr}_{1,2,..s}  \bigg(\frac{1}{{|{\bf r}'_{ij}|}} \cdot \frac{1}{({\bar h}^2 \tau_s - z)} \bigg)^n.
\label{eq_46}
\eea
The present structure consists 
of the sums of particle pairs connected by $\xi_{ij}$ interaction factors and should be considered as a series with respect to the powers $\xi_{ij}^n$. We may graphically represent all expressions by a sum of Mayer-like graphs consisting of points or bubbles (traces) and lines (interactions) which is quite similar to the $\lambda$-expansions analyzed in detail by Balescu \cite{Balescu}. The basic topological structure corrsponds to Mayers graph technique. However, the circles and lines we  use do not have exactly the same meaning as we speak regarding the Kelbg--Morita diagram  technique \cite{Kelbg62,Kelbg62a}.
Examples for the simplest Morita--Kelbg graphs are given in figure~\ref{Egraphs0}.  The terminus Mayer-like means here that the diagrams are structurally 
similar to the Mayer diagrams for the classical case. However, the wavy lines represent Kelbg interactions instead of Coulomb interactions 
and the lines representing  $\psi$-functions contain Slater functions instead of classical Boltzmann factors 
\cite{EbFoFi17,Fortov20}.
The triangles in the center of a graph stand for proper 3-particle non-additive interactions, which we approximate here 
by pair products.
We already see in the graph structure  the roots of the cropping methods developed later. Cropping means formally that in bond state configurations, 
low order contributions in terms of $e^2$ should not be included into partition functions. The other coefficients are defined by cluster integrals, e.g., \cite{Czerwon,BaEb71,EbHoKe67,EbHoKe67a,EbHoKe67b,EKK76,KKER86}.
The coefficients in terms of fugacity and density expansions are up to  a certain order exactly known \cite{Friedman,Czerwon,
Rogers86,Rogers01,KKER86}. 
In this quite complicated expansion, the coefficients $A_2$, $A_6$, $A_{10}$, $\dots$ represent the unscreened second and third virial coefficients
which we need to extract the mass action constants for the formation of pair and triple bound states.
Since all  higher orders $A_k$ with $k > 5$ are only approximately known, we already have to find for helium and lithium, the  appropriated approximations describing the binding effects.

\subsection{Identifying the free and the bound state contributions}

Assuming in the lowest order a Kelbg-like screening, we may write the
statistical thermodynamics in the form of cluster expansions in the canonical ensemble which also had already been discussed
in the preprint \cite{KelbgEb70,KelbgEb70a} and was worked out later \cite{KKER86,EbFoFi17,Fortov20}. The negative
free excess energy of classical and quasi-classical systems reads in the canonical ensemble \cite{Friedman,KKER86,EbFoFi17,Fortov20,
Brilliantov}
\bea
F = F_{\rm FD} + F_{\rm HF} + F_{\rm ring} +  \sum_{i} {\rm S}_{i}^{(k)} ,
\label{eq_47}
\eea
where the first 3 terms are the Fermi--Dirac, Hartree--Fock and Montroll--Ward contribution, which collects the contributions of ring diagrams \cite{EbKrRo12,EbKrRo12_a}.
These contribitons are studied more in detail elsewhere~\cite{KKER86,EbFoFi17}. 
The sums should be extended over the species of ions $i$ and also to all orders of clusters $k$. The cluster integral of 3rd order may be expressed by screened 2- and 3-particle Slater functions
\bea
{\tilde S}_{ij} = S_{ij} \exp(g_{ij} - \beta V_{ij}); \quad {\tilde S}_{ijk} = S_{ijk} \exp[(g_{ij} - \beta V_{ij}) + (g_{ik} - \beta V_{ik})+(g_{kj} - \beta V_{kj})].
\label{eq_48}
\eea
The final form of the 3-particle cluster integral including 3-particle quantum interactions is in the quantum case given by
\bea
S_i^{(3)} =  \frac{1}{2 \cdot 3} n_i \sum_{jk} n_j n_k \int \rd {\bf r}_j\, \rd {\bf r}_k \big({\tilde S}_{ijk}+ {\tilde S}_{ij} g_{jk} g_{kj} +
 g_{ij} {\tilde S}_{jk} g_{kj} + g_{ij} g_{jk} {\tilde S}_{ki} - 2 g_{ij} g_{jk} g_{ki} \big).
\eea
Here, the first term which also includes proper 3-particle interactions gives the largest terms at a strong binding 
and provides the asymptotics. Evaluating the integrals,  the cluster series have the following structure in the 
grand-canonical density representation
\begin{align}
\beta p(z_i, T) &= \beta p_{\rm FD} + \beta p_{\rm FD} +  \sum_i z_i Z_i^2 \bigg\{\sum_j \frac{2 \piup  \ell^2}{3} z_j  Z_j^2 {\tilde f}_i(K \lambda_{ie})
+ 2 \piup \sum_j z_j \lambda_{ij}^3 \big[ G_{ij}(\kappa) +  K_{20}(\xi_{ij}) \big]\nonumber\\
 &+ 8 \piup^2 \sum_{jk} z_j \lambda_{ij}^3 z_k \lambda_{ik}^3 \big[{\tilde G}_{ijk}(\kappa) +     {\tilde K}_{30}(\xi_{ij},\xi_{ik},\xi_{jk}) \big] + \ldots\bigg\}. 
\label{pdezsexp}
\end{align}
In a different form, we may represent the cluster expansion of the pressure within the canonical ensemble in terms of densities
\cite{KKER86,Alastuey95,Alastuey08,Alastuey08_a}
\begin{align}
\beta p(n_i, T) &= \beta p_{\rm FD} + \beta p_{\rm HF}  - \sum_i n_i \bigg\{ \frac{2 \piup \ell^2}{3 \kappa} Z_i^2 \sum_j n_j Z_j^2 \varphi_i (\kappa \lambda_{ij})
+2 \piup \sum_j n_j \lambda_{ij}^3 \big[G_{ij}(\kappa) +  K_{20}(\xi_{ij}) \big] \nonumber\\
 &+8 \piup^2 \sum_{ik} n_j \lambda_{ij}^3 n_k \lambda_{ik}^3 \big[G_{ijk}(\kappa) + K_{30}(\xi_{ij},\xi_{ik},\xi_{jk}) \big]
+ \ldots\bigg\} . \nonumber
\label{pdensexp}
\end{align}
Here, $\varphi(x)$ is in Debye--H\"uckel-like approximation $\varphi(x) = 1 - 3 \sqrt{\piup} / 8 + (3/10) x^2 + \ldots$. We do not specify the screening functions $G$, and  ${\tilde G}$, which depend on the screening parameters $\kappa$ (canonical) and $K$ (grand canonical ) which are
irrelevant in the present context of bound states and refer to \cite{KKER86}. 
We note that the graphs of Feynman-type like in figure~3 representating the  terms in the pressure expansion for hydrogen plasmas and for helium plasmas  which we should calculate contain at least one 4-ring ladder \cite{EbRoCPP24}.
Figure~\ref{qupress11} shows, as demonstrated already in 1970 at a seminar in Lviv, that fugacity cluster expansions are an appropriate instrument for the 
description of bound states without using any chemical tools such as mass action laws. A few years later this approach was taken up by the school of De Witt and Rogers in Livermore for the development of big astrophysical program packets ACTEX. Note that in his approach Rogers used numerical estimates of the cluster integrals equations~(26) and (41) in the fugacity expansion~\cite{Rogers,Rogers_a,Rogers79,Rogers79_b,Rogers86,Rogers01}. The program packet ACTEX (activity expansions) developed by Rogers has been used as one of the fundaments for the big project OPAL used for astrophysical calculations of the composition and opacity of stars and in particular for the
NASA Astrophysics Data Systems~(ADS)~\cite{Rogers,Rogers_a,Rogers79,Rogers79_b,Rogers01}. 
However, the OPAL equation of state is limited with respect to two aspects.
Firstly, it is only available in the form of pre-computed tables that are provided by
the Lawrence Livermore National Laboratory. However, these numerical applications require interpolations, which always lead to a loss of accuracy. 
Secondly, the OPAL equation of
state is proprietary and is not freely available. Our method is with respect to the theoretical tools comparable to OPAL. It is, however, completely analytical
and freely available~\cite{EbRoRe21,EbRoRe21_a}. Further we notice in this context that it was developed first in Lviv under the auspices of Ihor~R.~Yukhnovskii~\cite{KelbgEb70,KelbgEb70a}.
Here, we  use only analytical estimates in order to proceed to a better understanding and also give a comparison with the method of
mass action laws on the basis of the canonical ensemble.\\

\subsection{The partition functions}
We define here the  partition functions and the corresponding mass action functions as  the convergent part 
of the virial coefficients in the fugacity expansion. Following this idea in a recent work~\cite{EbRoCPP24} we developed 
a new method for calculating partition functions of atoms which is an alternative to 
the traditional chemical approach.
Our basic assumption is that the partition function of the atoms of hydrogen, helium and lithium can be represented by a convergent series with respect to the dimensionless reciprocal temperature 
\bea
\tilde \beta = {\rm Ry}/ k_{\rm B} T ,
\eea
resulting from the quantum statistical expressions for virial coefficients:
\bea
\sigma_{\rm atom} (T) = \sum_{s_0}^{\infty} A_{s_{0}} {\tilde \beta}^{s_{0}},
\label{Aseries}
\eea
with $s_0 = 2,4,6$ for hydrogen, helium, and lithium plasmas.

{\bf Hydrogen}:
The information about two-particle bound states is contained in the virial function  
$K_{20}(\xi)$, which can be split in a contribution of free and bound states. 
Note that in the present definition of the 2nd virial function, the linear and the quadratic terms are missing since they are already included into the Hartree--Fock and the Debye--H\"uckel term. 
Formally we get the bound state part first by splitting into the negative contributions of lower orders which are influenced by the screening effects and the positive contributions of the bound states \cite{KKER86}
\bea
K_{20}  (\xi) = - \frac{C_p }{12}\xi^3 +  \sqrt{\piup} \sum_{m \geqslant 4} \frac{\zeta(m-2)}{2^m \Gamma(m/2 +1)} \xi^m.
\eea
The bound state part of the 2nd virial function $K_{2b}$ is given by the even part of the virial function  
\bea
K_{2b}= \frac{1}{2}[K_{20} (\xi) + K_{20} (-\xi) ] = \sqrt{\piup} \sum_{m = 4,6,..} \frac{\zeta(m-2) \xi_{ij}^m}{2^m (m/2)!}= 2 \sqrt{\piup} \sum_{s = 1}^\infty \bigg[\exp\bigg(\frac{\xi^2}{4 s^2}\bigg) - 1 -  \frac{\xi^2}{4s^2}\bigg]. \nonumber 
\eea
The full function $K_{20}(\xi)$ may be decomposed into a bound and a free state part, alternatively we may split it into a direct and an exchange part \cite{KKER86}.
The point where these aymptotic representations are non-analytic is $\xi_{ij} = 0$, since here the forces change from positive to negative values
and at the same time the character of the symmetry changes. The quantum virial functions $Q (\xi)$ and $E(\xi)$  are discussed in detail in~\cite{KKER86,EbFoFi17,Fortov20}. The Taylor series for the quantum virial functions are convergent for any physical meaningful value of the interaction parameter $\xi$. This is ensured, since 
for large powers the terms show the same behaviour as the series of the well-known exponential function. 
This way the coefficients of our series are known and have the values
\bea
A_0 = A_1 = 0, \quad A_2 = 1.0967, \quad A_3 = 0.1078, \quad A_4 = 0.0056,\, \dots, \quad A_s = 4^s [\zeta(2s -2)/(2 s)!] .
\eea
The series are convergent since the $\zeta$ functions converge to one. This means that the coefficients converge to those of an exponential series.
We use the approximation
\bea
\sigma_H = 1.097 {\tilde \beta}^2 + 0.108 {\tilde \beta}^3 + 
\big[\exp({\tilde \beta}) - 1 - {\tilde \beta} - 0.5 {\tilde \beta}^2 - 0.166 {\tilde \beta}^3   \big].
\eea

The basic assumption of this work is that the partition functions of He and higher atoms follow the same scheme.
We evaluate the lowest coefficients as the binding parts of the cluster coefficients in quantum-statistical
fugacity expansions of the pressure, by evaluating the convergent graphs.
Based on the analysis of these virial coefficients, we identified the atomic partition functions and mass action constants as
convergent higher-order contributions to the unscreened virial coefficients.
The  most important selection criterion among the large number of graphs is that the contributions are strictly positive and increase strongly with $\beta = 1/ k_{\rm B} T$.
The problem with this method is that the corresponding higher order diagrams are difficult to
calculate since they include unknown three- and four-body quantum problems and further since the
number of relevant diagrams increases with the particle number.

Comparing the new partition function with the traditional approach of chemical physics to define the partition
functions as sums of terms
$\exp(-\beta \epsilon_n)$ where the $\epsilon_n$ are expressed by the known
energy levels of the atom, we find that the new partition functions are modified in the spirit of Planck--Brillouin--Larkin (PBL) by introducing the  elements of cropping.
In other words, the exponential functions are replaced by functions without the lowest powers
in ${\tilde \beta} = {\rm Ry} / k_{\rm B} T$. In this way, we retain the basic features and the low-temperature behavior of the traditional approach. However, on the other hand, we avoid any double counting of terms of lower order in $e^2$, which in Coulomb systems are used for purposes of screening  and electroneutrality and are no more free for the purposes of binding.

As already shown above, considering  \cite{EbRoMDPI,EbRoCPP24}, we have nowadays a complete understanding of the problem for hydrogen based on the  exact results from the quantum statistics of hydrogen \cite{EKK76}.
The complete series for the partition function of H are known, and we see that a low-order Rydberg expansion
obtained from quantum statistics works for $\tilde \beta \leqslant 2$ and agrees well with
the correct PBL partition function. A representation by two Taylor terms works already well at high temperatures with $\tilde \beta < 2$. On the other hand, the naive exponential approach agrees with the Planck partition for $\tilde \beta \geqslant 3$.
We find that the transition occurs in a rather interesting region covering for hydrogen temperatures around 75~000~K
where the binding effects are weak. The first orders in the Taylor expansion with respect to Ry correspond
to the so-called ladder diagrams of quantum statistical graph expansions (see figures~\ref{Mayer} and~\ref{Egraphs0}).
The quantum statistical analysis confirmed Planck's results from 1924 \cite{KKER86} as well as the findings of
Klimontovich for weak deviations from equilibrium \cite{Klim75}. 

\begin{figure}[htbp]
\begin{center}
\includegraphics[scale=0.4]{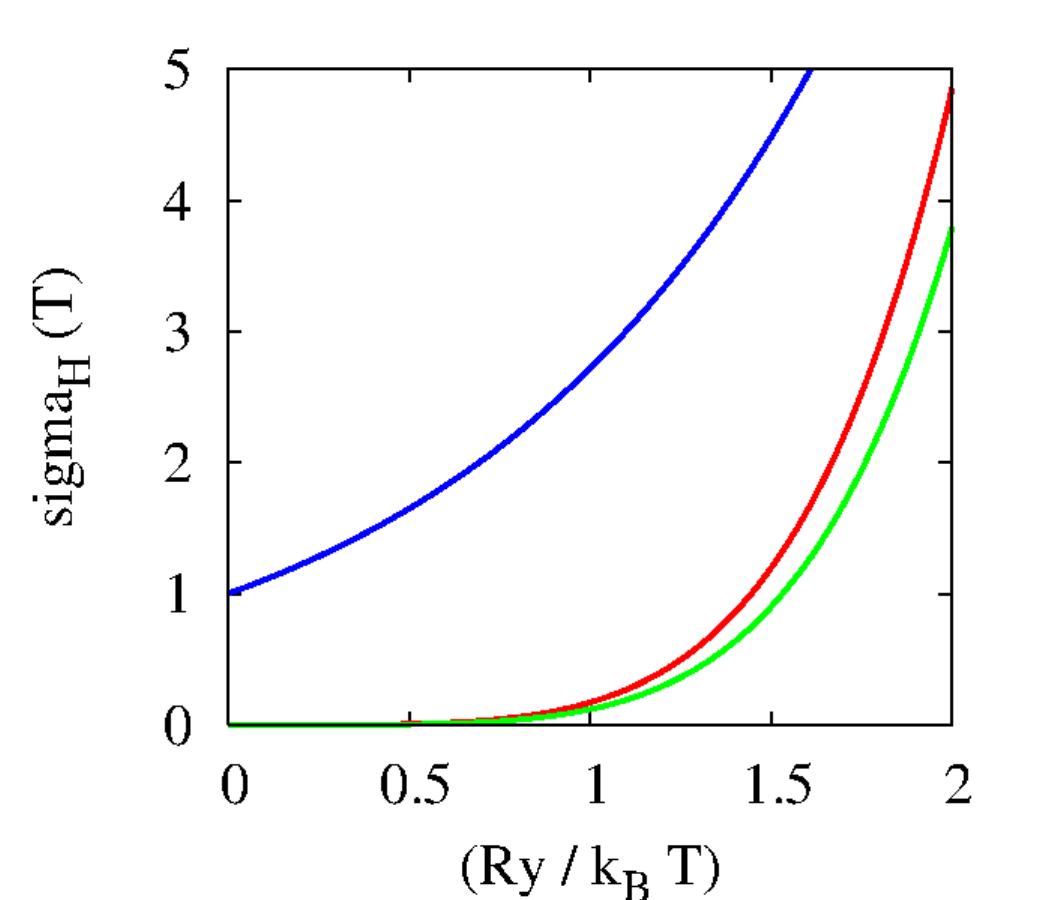}
\includegraphics[scale=0.4]{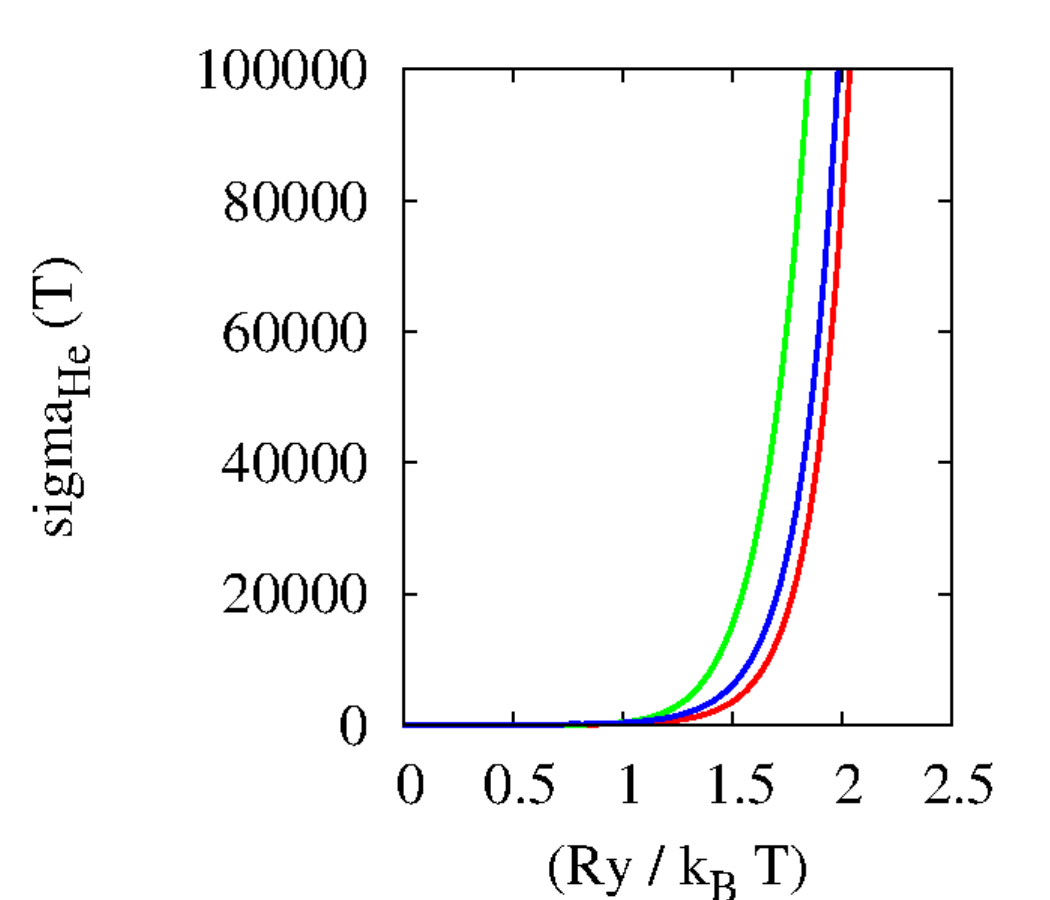}
\caption{(Colour online) Partition function of H (left-hand) and He (right-hand). The green curves show the first terms of a Taylor expansion, the blue curves show the exponential approximation and the 
red curves interpolate between the Taylor series and the asymptotic exponential representation.}
\label{partf}
\end{center}
\end{figure}

We consider now the partition functions for
{\bf helium}:
Here, the calculations are not yet fully finished. The results given in \cite{EbRoMDPI,EbRoCPP24} based on diagrammatic calculations of the lowest binding diagrams
including ireducible diagrams (shown in figures~\ref{Mayer} and~\ref{Egraphs0}) as well as reducible diagrams (shown in figure~\ref{Egraphs0})
provided the estimate \cite{EbRoMDPI,EbRoCPP24}
\bea
\sigma_{{\rm He}} \simeq 14.90 {\tilde \beta}^4 + 373  {\tilde \beta}^5 .  
\label{sigmaHe3}
\eea
On the other hand, the asymptotics are determined by the ground state of the helium atoms ${\tilde \epsilon}_{\rm He} \simeq 5.087$
and we find:
\bea
\sigma_{\rm He}  \sim  \exp({\tilde \beta} {\tilde \epsilon}_{\rm He}) .
\eea
We combine both results and propose to take the lower order terms from the evaluation of the graphs and the higher order terms from the asymptotics to find \cite{EbRoMDPI,EbRoCPP24}
\bea
\sigma_{\rm He}  =   14.90 {\tilde \beta}^4  + 373 {\tilde \beta}^5+ \big[\exp({\tilde \beta} {\tilde \epsilon}_{\rm He} )\big]' 
=  14.90 {\tilde \beta}^4 + 373 {\tilde \beta}^5 + \sum_{k=6}^{\infty}
\frac{( {\tilde \epsilon}_{\rm He} {\tilde \beta})^k}{k!} .
\label{sigmaHe5}
\eea
On the other hand, we have a different estimate based on the tables of the excitation energies of helium taken from 
\cite{NIST,EbRoMDPI} as well as on the theoretical insight, that cropping should be applied to each of the electrons extra.
These representations suggest that Taylor expansion in the dimensionless ${\tilde \beta}$-parameter as well as the asymptotics start for helium with the terms \cite{EbRoMDPI}: 
\bea
\big[\exp(4 {\tilde \beta}) - 1 - 4 {\tilde \beta} \big] \cdot \big[\exp(1.807 {\tilde \beta})-1-1.807 {\tilde \beta} \big].
\eea
Avoiding any overlap of the lower orders, we find another estimate 
\bea
\sigma_{\rm He} (T) \simeq 14.90 {\tilde \beta}^4 + 373 {\tilde \beta}^5 + \big[ \exp(4 {\tilde \beta}) - 1 - 4 {\tilde \beta}  - 8 {\tilde \beta}^2 \big] \\ \nonumber
\times \big[\exp(1.807 {\tilde \beta})-1-1.807 {\tilde \beta} -  1.622  {\tilde \beta}^2 \big].  
\eea
This gives, e.g., $\sigma_{\rm He} (100\;000~\textrm{K}) \simeq 20464$. We have applied here extra cropping steps to the exponent of each of the electrons to obtain the terms of the order $\tilde \beta^n$ with $n \geqslant 6$
which complement our diagrammatic calculations. To obtain better results, we need to calculate more diagrams of the order $\tilde \beta^n$
with $n = 5,6, \ldots$.
In figure~\ref{partf}, we compare  the low order expansions for hydrogen and for helium with approximations, that interpolate between a power law and exponential functions.
Note that the corresponding mass action function is
\begin{align}
K_{\rm He} (T)  &= (2 \sqrt{\piup}  \lambda_{ie}^3)^2  \Big\{ 14.90 \tilde \beta^4  + 373 \tilde \beta^5  \nonumber\\
&+ \big[ \exp(4 {\tilde \beta}) - 1 - {\tilde \beta}  - 8 {\tilde \beta}^2 \big] \big[\exp({1.807 {\tilde \beta}})-1-1.807 {\tilde \beta} -  1.622 {\tilde  \beta}^2 \big]\big\}.
\end{align}

The new results
are generally much lower than the naive exponential approach without cropping.
Due to these differences, we recommend for further applications the formulae which interpolate between the Taylor expansions and the asymptotics, i.e., between higher and lower temperatures. The interpolating formulae are shown as red curves. For lower temperatures, the new formulae are practically identical to the asymptotics and for higher temperatures to the estimates calculated from the diagram expansions.
We can see for hydrogen that the low-order diagrams are only first approximations and should be supplemented by calculating further diagrams. 
We show later that the at high temperatures the concentrations, predicted for the atom formation, are lower than in previous calculations. This can be important for estimating ionization in stars and the fusion devices, for example.

From the point of view of our cluster analysis, the consideration of bound states requires the following steps:
\begin{enumerate}[label=(\roman*)]
	\item Identify the convergent contributions of two-, and three- particle clusters to the coefficients of the fugacity or density expansions.
	\item  Identify the positive definite and, at low $T$, asymptotically large contributions that define the mass action constants.
	Note that the candidates for the contributions of bound states are diagrams with an even number of lines. They should contain at least 
	one 4-fold $(+-)$ interaction line like in figure~\ref{Mayer}.
	\item Most important: quantum statistics of Coulomb systems teaches us that a class of diagrams which are, e.g., of linear order in $e^2$ or are of ring type are consumed by processes of screening or cancelled out by electro-neutrality. Such diagrams should not be included into the binding expressions like mass action functions. This  is the origin of the cropping steps, which we applied. Following Onsager's postulates, any double-counting of contributions should be strictly avoided. A graph can either 
	provide a contribution of free states or of the bound states.
\end{enumerate}

\section{The equation of state for hydrogen plasmas}

Analyzing  equation~(\ref{pdezsexp}) 
we find the structure
\bea
\beta p(n_i, T) = \beta p_{\rm FD} + \beta p_{\rm HF}   + \beta p_{\rm sc} + \beta p_{\rm bo},
\eea
with a bound state (discrete state) contribution
\bea
\beta p_{\rm bo} = +2 \piup \sum_j n_j \bigg\{ \lambda_{ij}^3 \big[K_{2b}(\xi_{ij}) \big] + 8 \piup^2 \sum_{ik} n_j \lambda_{ij}^3 n_k \lambda_{ik}^3 \big[K_{3b}(\xi_{ij},\xi_{ik},\xi_{jk}) \big] 
+ \ldots\bigg\} , \nonumber
\eea
and a contribution of scattering states (free states)
\begin{align}
\beta p_{\rm sc} &= - \sum_i n_i \bigg\{ \frac{2 \piup \ell^2}{3 \kappa} \sum_j n_j Z_i^2 Z_j^2 \varphi_i (\kappa \lambda_{ij})
+2 \piup \sum_j n_j \lambda_{ij}^3 \big[{\tilde G_{ij}}(\kappa) +  K_{2f}(\xi_{ij}) \big]  \nonumber\\
 &+8 \piup^2 \sum_{ik} n_j \lambda_{ij}^3 n_k \lambda_{ik}^3 \big[{\tilde G_{ijk}}(\kappa) +     K_{3f}(\xi_{ij},\xi_{ik},\xi_{jk})\big] 
+ \ldots\bigg\} . \nonumber
\end{align}
Let us now consider the simplest case of low-density hydrogen plasmas. 
Based on th present analysis, the EoS for hydrogen plasmas reads in the lowest order in the density, i.e., including only quadratic terms in the density \cite{KKER86,EKK76}
($n_i$-density of the nuclei):
\begin{align}
\beta p  &= 2 n_i   - A_0 (T) n_i^{3/2} - n_i^2 B_2 (T) + \ldots \, , 
\label{eq_63}\\
A_0(T) &= (\sqrt{8 \piup} / 3) \ell^{3/2}; \quad B_2 (T) = \big[8\piup \sqrt{\piup} \lambda_{ie}^3 \sigma_{PBL} (T )   + K^*(T) \big].
\label{eq_64}
\end{align}
Here, the bound state contribitons are the PBL-partition function and the contributions of scattering states in the 
correction function $K^*(T)$ which is numerically relatively small. Larkin has shown that in the classical case $K^*(T) =0$.  
A simple low-temperature approximation is 
\bea
K^*(T) \simeq - \sqrt{\piup} \lambda_{ie}^3 \xi_{ie}^2 .
\label{eq_65}
\eea
We may use this expression to define a quantum length (Kelbg length)
by
\bea
K^*(T) = - 4 \piup \ell^2 a^K(T) .
\label{eq_66}
\eea
The Kelbg length $a^K(T)$ is a kind of quantum Debye--H\"uckel distance between the charges.
The full quantum function $K^*(T) $ is exactly known; it can be expressed by the 
quantum virial functions defined above \cite{KKER86,EKK76}. The functions are also tabulated \cite{EKK76}. 
This way, the quantum statistical analysis confirms Planck's results from 1924 \cite{EKK76} as well as the findings of 
Klimontovich \cite{Klim75}. Evidently there are no open problems left for hydrogen.

For a more convenient way of calculating the free densities of electrons, ions (protons) and atoms $n_e^*$, $n_i^*$, $n_0^*$,
we go to the chemical picture introducing free and bound particles. 
\bea
\beta p  = n_i^* + n_e^*  + n_0^*   - \frac{\kappa^{*3}}{24 \piup} \Big[1 - \frac{3}{2} \kappa^* a^K(T) + \ldots \Big] , \qquad
n_0^* = n_e^* n_i^* 8\piup \sqrt{\piup} \lambda_{ie}^3 \sigma_{PBL} (T ).
\eea
Here, $\kappa^{*2} = 4 \piup \ell (n_e^* + n_i^*)$ defines the Debye quantity of the free particles, and the conservation relations are 
$n_e = n_e^* + n_H^*$, $n_i = n_i^* + n_H^*$. 

\begin{figure}[htbp]
\begin{center}
\includegraphics[scale=0.35]{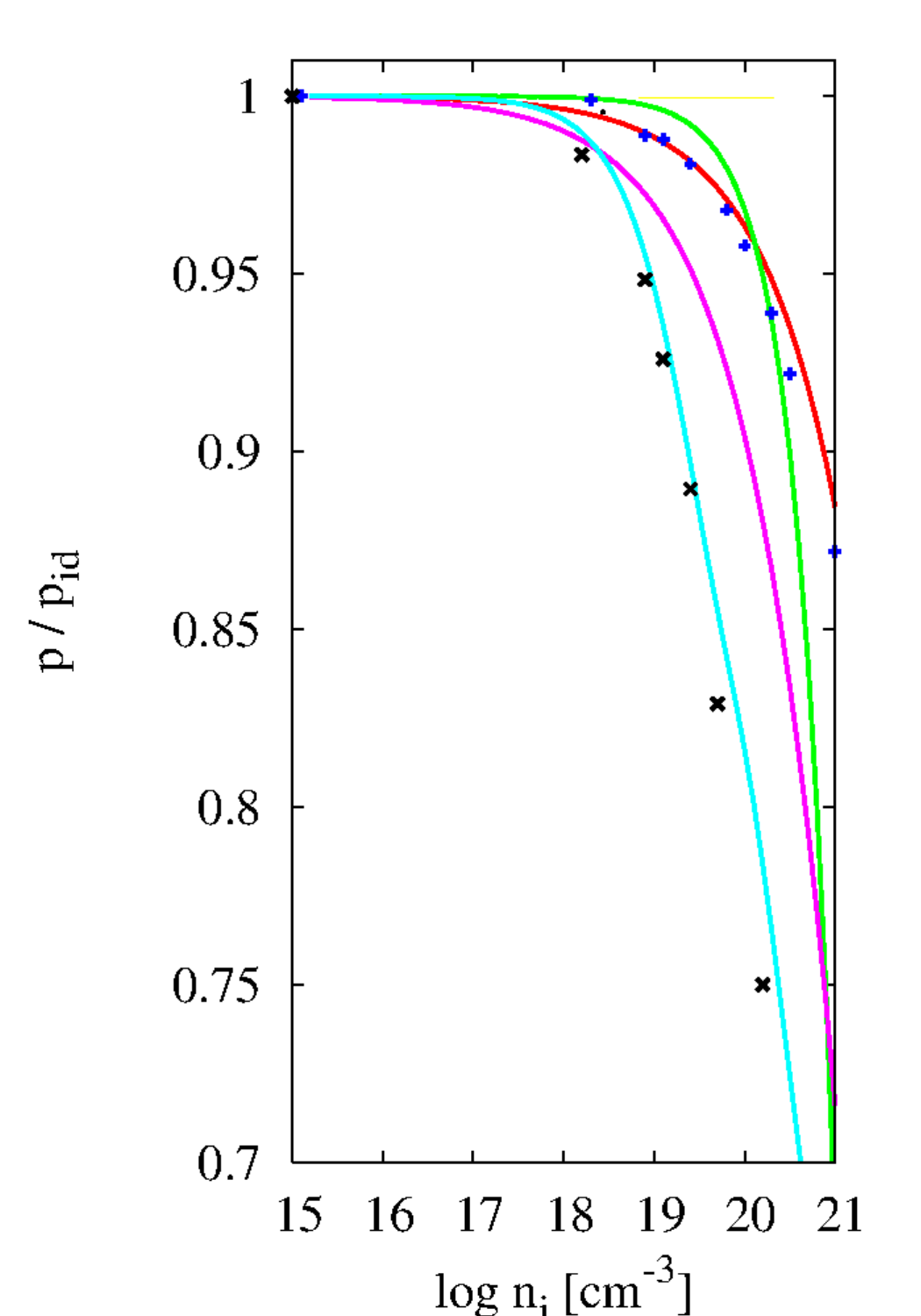}
\caption{(Colour online) The EoS of hydrogen calculated within the Kelbg--Yukhnovskii approximation (QKYA) for the temperatures $T = 31250$ K (turquoise line) and $62500$ K  (red
line), and a comparison with  data from recent Monte Carlo PIMC calculations \cite{BonitzPIMC,BonitzPIMC_a}. The two additional lines (magenta and green) show the density expansion including only the terms up to the
quadratic order after equations~(\ref{eq_64}--\ref{eq_65}).}
\label{pgraph1}
\end{center}
\end{figure}

 The Kelbg length is an aggregated quantity which includes several contributions from the scattering states.
Using the analogies between the classical 
Debye--H\"uckel theory and this quantum approach to plasmas we get as in \cite{EKK76,EbFoFi17} the so-called quantum Debye--H\"uckel approximation QDHA which was obtained in \cite{EKK76,EbFoFi17} by summing up further higher order terms
in the density  
\bea
 p  = n_i^* + n_e^*  + n_0^*   - \frac{\kappa^{*3} \varphi(\kappa^* a(T))}{24 \piup}; \quad \kappa^{*2} = 4 \piup \ell (n_e^* + n_i^*);\\
 \frac{n_0^*}{n_e^* n_i^*} = 8\piup \sqrt{\piup} \lambda_{ie}^3 \sigma_{PBL} (T ) \exp\big[- \kappa^* \ell G(\kappa^* a(T))\big];\\
G_{\rm DH} = \frac{1}{1+x}; \qquad  \varphi_{\rm DH}(x) = 1 - \frac{3}{2} x + \frac{9}{5} x^2 - 2 x^3 + \ldots 
\eea
Note that this convenient approximation which is often used for hydrogen calculations  of free densities~\cite{EKK76,EbFoFi17} is at low densities equivalent to the 
exact equations~(\ref{eq_63}--\ref{eq_66}) and differs only in the approximative higher orders.  An essential insight is that the correction to the PBL-expression $K^*(T)$ is not a contribution to binding but stems from the  scattering states and should be treated together with the Debye term~\cite{EKK76,KrScKr05}.
This is possible, since the contributions depend only on $\kappa^*$, the Debye parameter of the free plasma. 
The QDHA is based on the analogies and an exptrapolation \cite{EbKrRo12,EbKrRo12_a}. Now we derive a better founded approximation within the schema of the exponential potential. This way we proceed from the DH-approximation to the new KY (Kelbg--Yukhnovskii) approximation. Following Kelbg, we define the quantities 
\bea
\frac{1}{\alpha^K (T)} = \frac{2}{3} a^K (T) = - \frac{2 K^*(T)}{12 \piup \ell^2} \simeq  \frac{\sqrt{\piup}}{6} \lambda_{ie}
\eea
and a  dimensionless quantum Kelbg parameter and a quantum Kelbg distance 
 \bea
 x  = q_K  = \kappa a^K (T) = (\sqrt{\piup}/4) \kappa \lambda_{\pm}; \qquad a_{ij}^K = (\sqrt{\piup}/4) \lambda_{ij}.
\eea
We define further  the Kelbg--Yukhnovskii  functions with $X = \sqrt{1 + 2 \kappa/ \alpha^K} = \sqrt{1 + (4/3)  \kappa a^K (T)} $ in analogy to the Debye--H\"uckel functions:
\bea
G_{\rm KY} (x) = \frac{6}{\sqrt{\piup} x} \bigg[ 1 - \frac{1}{\sqrt{1 + (\sqrt{\piup}/3) x}} \bigg] = 1 - \frac{\sqrt{\piup}}{4}x + \ldots\, ,\\
\tau_{\rm KY}(x) = \frac{1}{x^3} \bigg( \frac{1}{4} X^3 - \frac{3}{2}X +2 - \frac{3}{4X} \bigg) = 1 -  \frac{3 \sqrt{\piup}}{8}  x + \ldots\,.
\eea 
This way we get the new formulae for the EoS of hydrogen plasmas denoted as Kelbg--Yukhnovskii approximation:
\bea
 p  = n_i^* + n_e^*  + n_0^*   - \frac{\kappa^{*3}}{24 \piup} \varphi_{\rm KY} (\kappa^* a^K (T)) ; \qquad \kappa^{*2} = 4 \piup \ell (n_e^* + n_i^*),  
\eea
with the mass action law 
\bea
\frac{n_0^*}{n_e^* n_i^*} =  8\piup \sqrt{\piup} \lambda_{ie}^3 \sigma_{PBL} (T ) \exp \big[- Z_i \ell \kappa^* G_{\rm KY} (\kappa^* a^K (T) ) \big].
\eea
Graphical representations show that the QDHA and the QKYA are similar in shape, although the QKYA is characterized by a longer range of the correlations as the QDHA and as the ring functions derived in~\cite{EbKrRo12,EbKrRo12_a}.
The  transition from the earlier ring functions QDHA \cite{EbKrRo12,EbKrRo12_a} to the present QKYA is in our view an improvement, since the Kelbg--Yuknovskii  functions are stemming from a systematical evaluation of ring diagrams and are better founded. We calculated with the new KY  approximations the pressure of an hydrogen plasmas and show the results for the two temperatures
 $T= 31 250$~K and  $T= 63 500$~K in comparison with the recent results from Path Integral Monte Carlo (PIMC) calculations. We see in figure~\ref{pgraph1} that the agreement is rather good up to $n_i = 10^{21}$~cm$^{-3}$.
 
We worked so far with just one quantum parameter, namely  $\kappa^* a^K (T)$. This gives already quite satisfactory results. 
We  show now that the method of averaging at the level of screening functions for pairs leads to better approxinations.
Let us start with the excess part of the chemical potential which is determined by the ring contributions. 
In extended QDHA, we have
\bea
G_{ij} = \frac{e_i^2 e_j^2 } {\big( \sum_i \sum_j  e_i^2 e_j^2 \big)} \frac{1}{1 + \kappa^* a_{ij}^K} .
\eea
The KY-theory gives the mass action law 
\bea
\frac{n_0^*}{n_e^* n_i^*} =  8\piup \sqrt{\piup} \lambda_{ie}^3 \sigma_{PBL} (T ) 
\exp \big[- Z_e Z_i \ell \kappa^* G(\kappa^* a^K_{\pm}) \big].
\eea
A graphical representation shows that the approximations QDHA and QKYA are similar in shape, although the QKYA based on the true ring functions for the Kelbg potential is characterized by a longer range of correlations as the QDHA which was based on the ring functions derived in \cite{EbKrRo12,EbKrRo12_a}.
We may represent the new expressions for $\tau$ and $G$ as functions of the quantum length $a^K = 3/2 \alpha^K$ and find
\begin{align}
G_{\rm KY} (\kappa a) &=  \frac{3}{2 a} \bigg( 1 - \frac{1}{\sqrt{1 + 4 a/3}}\bigg)  =1 -  (\kappa a) +\frac{10}{9}(\kappa a)^2  - \frac{35}{27}  (\kappa a)^3 + \ldots\,, \\
\varphi_{\rm KY} (\kappa a) &= 1 - \frac{3}{2} (\kappa a) + 2  (\kappa a)^2 - \frac{70}{27} (\kappa a)^3 - \ldots\, .
\end{align}
Note that the DH and the KY theories are similar in shape but have different asymptotics
\bea
G_{\rm KY} \sim \frac{3}{2 \kappa a}; \quad  G_{\rm DH} \sim \frac{1}{\kappa  a}; \quad \varphi_{\rm KY} \sim \frac{2}{3 \sqrt{3} (\kappa a)^{3/2}}; \quad  \varphi_{\rm DH} \sim \frac{3}{(\kappa  a)^2} .
\eea
The excess part of the chemical potentials which determines the Saha  equation is now given by 
\bea
\mu_i^{\rm ex} = \mu_i^{\rm ring} = - k_{\rm B} T Z_i \ell \kappa^* \sum_j G_{ij}(\kappa^* a_{ij}^K ; n_i).
\eea
An explicite form is
\bea
\mu_i^{\rm ex} = - \frac{3 e_i^2 \alpha}{4 \piup \epsilon_0 \epsilon_r k_{\rm B} T 2 a^K \kappa^2} \bigg[ 1 - \frac{1}{\sqrt{1 + (4/3)  (\kappa a^K)}}  \bigg].
\eea
In the same way we may calculate the Coulomb energy.

\section{The equation of state for helium plasmas}

We consider the example of hydrogen as a good plan how to proceed for helium and begin with the virial expansion in the density. 
An extension of the virial series for  $Z > 1$  elements reads with the notation $n  = n_e + n _i = (1 + Z) n_i$, where $n_i$ is the density of the nuclei~\cite{KKER86}:
\begin{align}
\beta p  = n   &- A_0 (T) n^{3/2} - n^2 B_2 (T) - \frac{3}{2}  A_3(T) n^{3/2} \ln n - A_4 (T) n^{5/2}  \nonumber\\
 &- 2 A_5 (T) n^3 \ln n  -  B_3(T) n^3 + \ldots
\end{align}
The concrete values of the temperature functions $A_i(T)$, $B_k(T)$ follow by a comparison with the cluster expansion and are given for $i = 1,2,3,\ldots$,
$k = 2,3$, e.g., in \cite{Czerwon,KKER86}. Our basic idea is that the basic convergent part of the 3rd virial coefficient
$B_3(T)$ can be identified with the partition function and the mass action function of the helium formation. The problem is
that the coefficients  $B_3 (T)$ and the coefficients beyond are not well studied yet. Therefore, the partition functions are 
often approximated by elementary expressions based on the known energy levels \cite{NIST}. Often these calculations neglect so far the compensation effects between the high discrete and the low scattering states except, e.g., in \cite{EbRoCPP24}. 
We know from the studies for hydrogen, that for $T > I/k_{\rm B}$, which is the ionization temperature, 
the way of cropping is essential and the compensation effects determine the value of the partition functions. We should expect that for helium 
the compensation effects between
high discrete and low scattering effects play an essential role, too. We must expect that these compensation effects dominate 
beyond the ionization temperatures, i.e., for helium about $I/k_{\rm B} \simeq 630 000$ K.
Let us consider first a representation for high temperatures where the bound states are irrelevant. Note that the contributions of bound states start 
with the power $\xi_{ie}^4$. The first terms of the pressure read
\begin{align}
 \beta p(n_i, T) =& \sum_i n_i  -   \frac{2 \piup \ell^2}{3 \kappa}  \sum_{ij} n_i n_j Z_i^2 Z_j^2 \varphi_i (\kappa \lambda_{ij})
   -  \sum_{ij} n_i n_j Z_i^3 Z_j^3 \ln \bigg[\frac{\kappa}{\alpha_{ij}}\bigg] \\ \nonumber
 &- 2 \piup \sum_j n_j \lambda_{ij}^3 K_{20}''(\xi_{ij})+ \ldots
\end{align}
The bound state contributions are  defined only for $\xi_{ij} > 0$. We  assume that the 3rd virial functions have the same structure as the 2nd virial coefficients, having a bound state and a free state contribution
\bea
K_{30} (\xi)  =  K_{3b}(\xi) + K_{3f}(\xi) .
\eea
We identify the contribution of the bound states to the 3rd virial function following the lines developed above for the partition function
of hydrogen.

\begin{figure}[htbp]
\begin{center}
\includegraphics[scale=0.47]{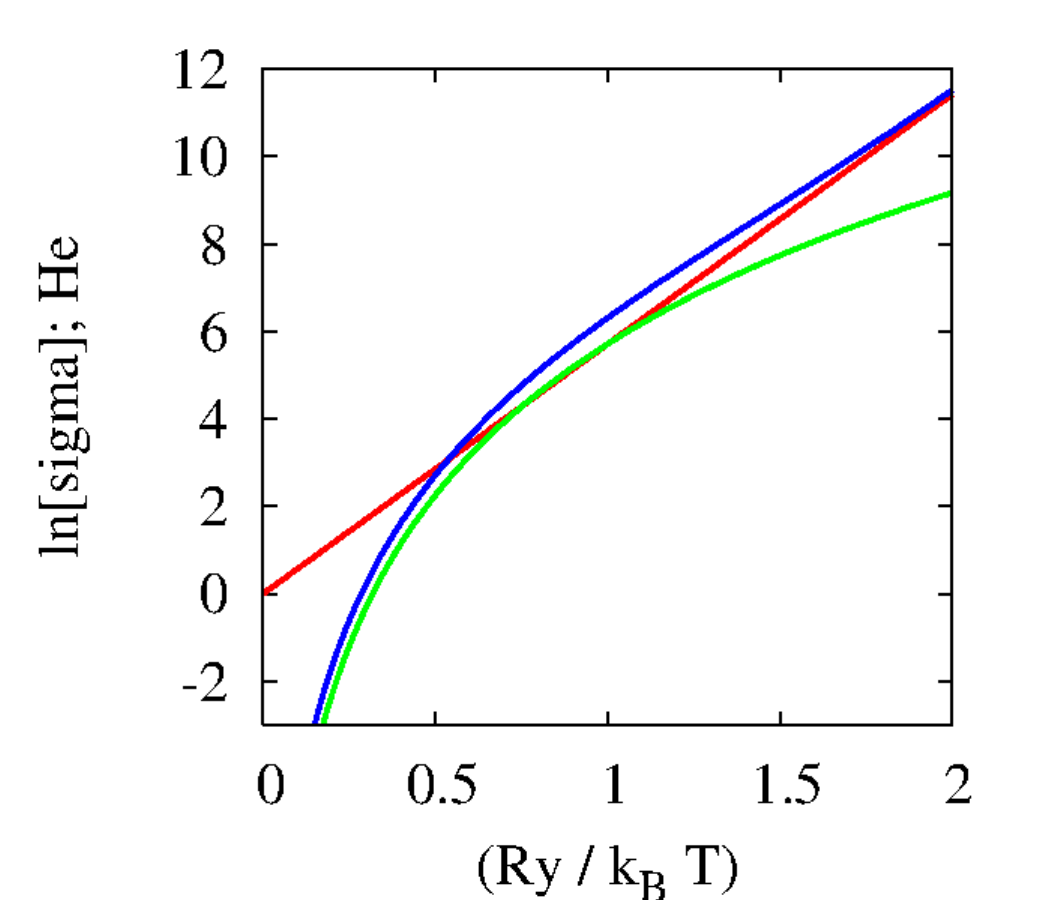}
\includegraphics[scale=0.29]{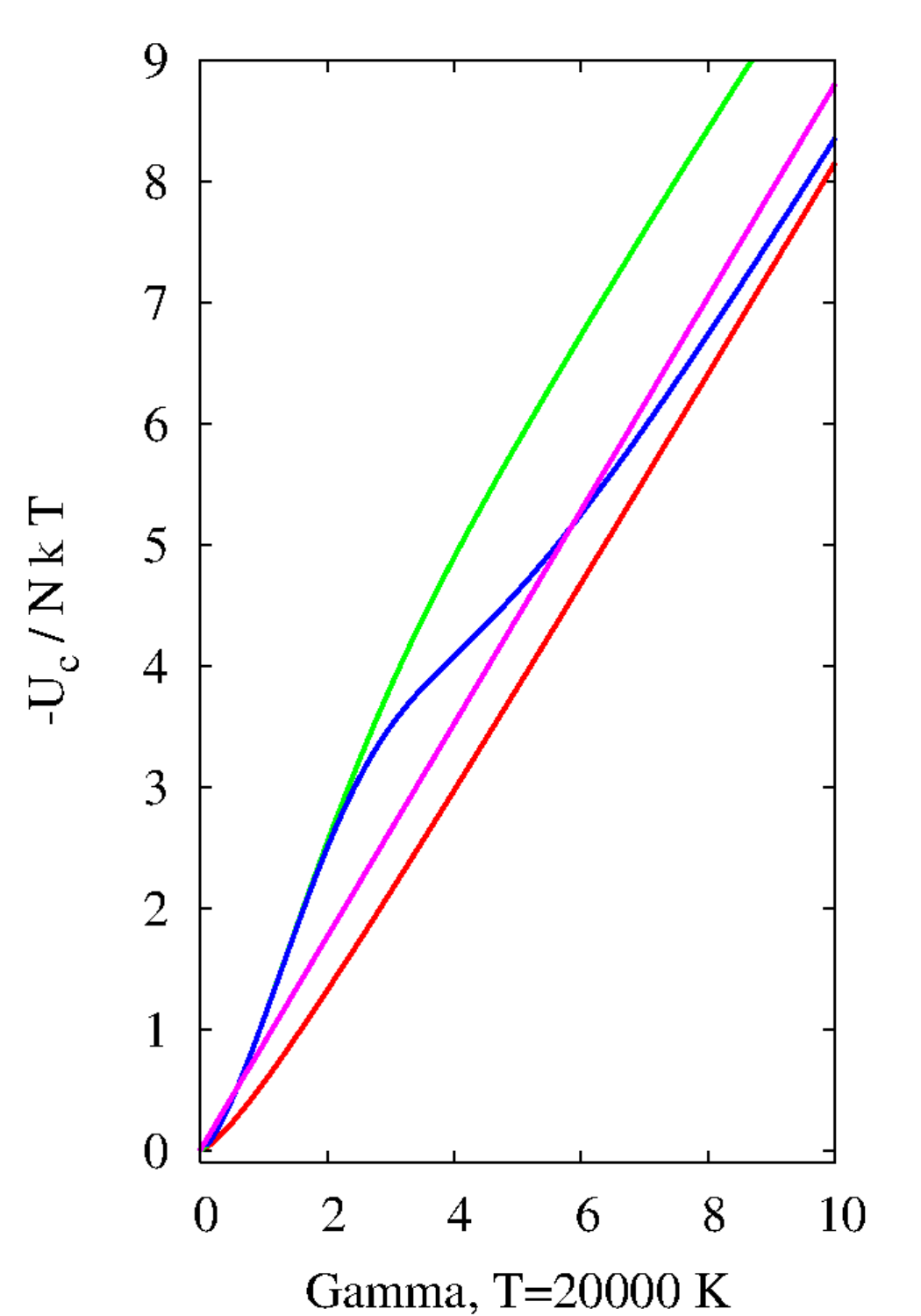}
\caption{(Colour online) Left-hand: comparison of several estimates for the partition functions
of helium. The relevant new results are for high temperatures in the left-hand lower corners, where the new results (blue curves) deviate from the traditional exponential partition functions (red curves).
The new (blue) curves were obtained by 
interpolations between a low order polynomial (green line) and the high order exponential term (red line).
Right-hand: comparison of the present new approximation for the Coulomb energy (blue line) with the Uns\"old aymptotics (magenta line), the estimate by Brilliantov (red line) and the Debye energy (green line) which is correct mainly for small densities.}
\label{partHeLi}
\end{center}
\end{figure}

We calculate now the EoS of helium plasmas and start with an ideal approximation.
For  He-plasmas we have the following  relations between the species
including mass action laws:
\begin{eqnarray}
 n_e &=& n_e^*+ n_{\rm He^+}^*  + 2  n_{\rm He}^* , \quad n_i= n_{\rm He^{++}}^*+n_{\rm He^{+}}^*+n_{\rm He}^*, \nonumber \\
n^*_{\rm He^+}&=&n_e^* n_{\rm He^{++}}^*K_{\rm He^+}(T), \qquad n^*_{\rm He}={n_e^*}^2 n_{{\rm He}^{++}}^* K_{\rm He}(T).
\label{compositionHe}
\end{eqnarray}
Charge neutrality means $n_e=Z n_i = 2 n_i$ so that the properties of the He plasma are determined only by $T$ and the total density of nuclei 
$n_i$. The ionization degree may be introduced as 
the relation between free electron density and total  electron density $\alpha_e = n_e^*/Z n_i$.
For the pressure, we have within the approximation of ideal particles 
\bea
\beta p = n_i^* (1 + Z)  + n_{\rm He^{+}}^* + n_{\rm He}^* .
\eea
Here, the density $n_i$ and the temperatute $T$ are the relevant independent variables and $n_j^*$, $n_{\rm He^{+}}^*$, $n_{\rm He}^*$ are dependent variables. We go now to a chemical description including nonideality effects, 
which follows the lines of our fugacity expansion.
Using this analogy we get the new formulae for the EoS of helium plasmas $Z=2$ with $n_i =n_{\rm He^{++}}^*$: 
\bea
 p  = n_i^*  + n_e^*  + n_{\rm He^{+}}^*  + n_{\rm He}^*   
   -   \frac{2 \piup \ell^2}{3 \kappa^*}  \big[Z_i^4 n_i^* \varphi (\kappa^* \lambda_{ii}) + n_e^{*2} \varphi (\kappa^* \lambda_{ee})\\
+ n_e^* n^*_{\rm He^+} \varphi (\kappa^* \lambda_{e,\rm He^{+}}) + Z_i^2 n_i^* n_{\rm He^+} \varphi (\kappa^* \lambda_{i,\rm He^{+}}) + Z_i^2  n_i^* n_e^* \varphi (\kappa^* \lambda_{ie}) \big]
  \eea
with the  sccreening parameter defined by the concentrations of all charged species
\bea
(\kappa^*)^2 = 4 \piup \ell (n_e^* + n_{\rm He^{+}}^* + 4 n_{\rm He^{++}}^*).
\eea
Following the line of our fugacity expansion, we get now two mass action laws 
\bea
\frac{n_{\rm He^{+}}^*}{n_e^* n_i^*} =  8\piup \sqrt{\piup} \lambda_{ie}^3 \sigma_{\rm He^{+}}^{\rm PBL} (T ) \exp \big[- \ell \kappa^* G_{\rm KY} (\kappa^* a_{e,i+}^K (T) ) \big],
\eea
and
\bea
\frac{n_{\rm He}^*}{(n_e^*)^2 n_i^*} =  8\piup \sqrt{\piup} \lambda_{ie}^3 \sigma_{\rm He} (T ) \exp \big[- 2 \ell \kappa^* G_{\rm KY} (\kappa^* a_{ie}^K (T) ) \big].
\eea
Here, the $a_{ij}^K (T) $ are appropriately chosen Kelbg lengths. We underline that the main source for this derivation is the comparision with
the exact fugacity cluster expansion. We have ensured that the chemical representation using the species $n_i^*$, $n_e^*$, $n_{\rm He^{+}}^*$, $n_{\rm He}^*$
is at low densities fully equivalent to the exact fugacity expansion and this way also to the exact density expansion.
A direct comparison of the partition functions and the EoS for He with other sources is difficult since in most applications, e.g., to stellar
athmospheres (see, e.g., \cite{Cardona10}) a different normalization is used. The mentioned authors follow in most cases the traditional definitions of Eggert, Saha and Uns\"old. Relative partition functions are defined for each ionization step $k$ and are denoted by $U_k(T, n_e)$. The relative partition functions 
depend on temperature and on densities, 
corresponding to density-dependent cutting of the summations over atomic states. Therefore, it may be easier to compare the 
particle densities predicted by  different methods.   
Due to our predicion of lower 
values of our mass action constants, we find at high $T$ smaller densities of free charged particles as in standard calculations.
In other words, corresponding to figures~\ref{partf} and~\ref{partHeLi}, we predict at higher temperatures smaller
atom densities than in the standard approaches \cite{Cardona10}. This provides us with the basic part of the EoS of helium.

\begin{figure}[htbp]
\begin{center}
\includegraphics[scale=0.3]{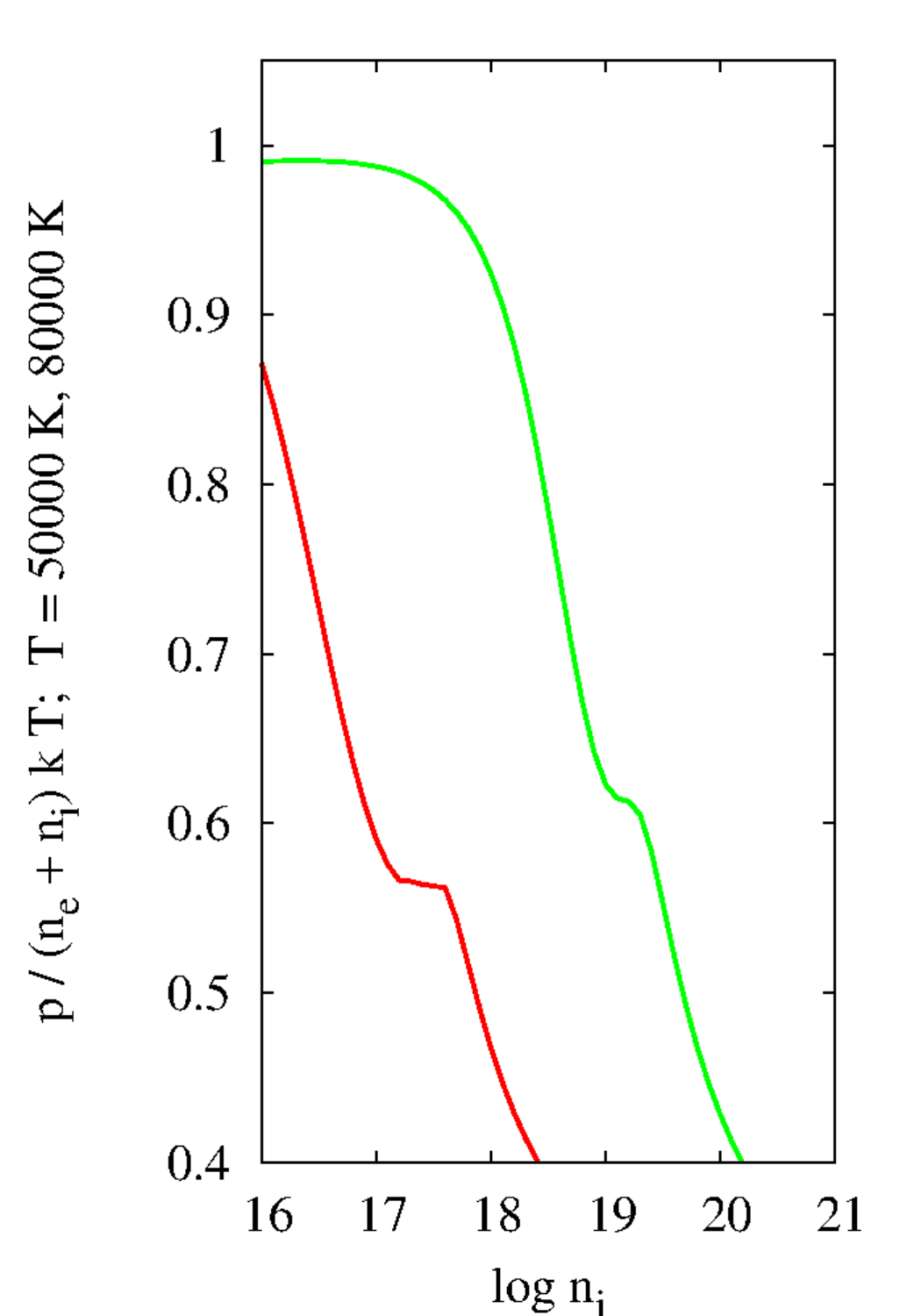}
\end{center}
 \caption{(Colour online) 
EoS of helium plasmas from the chemical approximation given above represented in the form of  the relation of total pressure to the ideal Boltzmann pressure at  densities
$n_i < 10^{21}\,{\rm cm}^{-3}$. We show the relative pressure $\beta p / (n_e + n_i)$  for $T = 50000, 80000$  K
(red, green curves) as a function
 of $\log (n_i)$. The shoulders correspond to the formation
of helium ions.}
\label{Hepress25}
\end{figure}

 In the last part of this chapter we want to show how our theory can be made 
compatible with Uns\"old's work on astrophysical plasmas. As shown in  \cite{Stolzmann,Brilliantov,Eb16,Eb16_a}, we find for $Z \gg 1$ and strong 
ion-ion correlations for the Coulomb energy and the corresponding lowering of ionization energy the asymptotics: 
\bea
U_{i,C} =  -a (N k_{\rm B} T) \Gamma; \qquad p_{i,C} = - \frac{a}{3} (n_i k_{\rm B} T) \Gamma_i .
\eea
There exist several estimates for the coefficient   \cite{Stolzmann}. The theory of Berlin and Montroll provides 
$a = (3/2)^{4/3} = 1.7171$, Uns\"old finds $a = 1.7861$ and Brilliantov $ a = 0.881$ \cite{EbFoFi17,Unsold,Stolzmann}.
We see that the slopes predicted by Berlin, Montroll and Uns\"old are in resonable agreement. On the other hand, the estimate by Brilliantov agrees well with a numerical fit of simulation data for classical plasmas by De Witt which provide $a = 0.899$  
and what may be even more convincing, with the bound given by Onsager in 1937 $a = 0.9$~\cite{EbFoFi17}. 
Sinnce the differences are not well understood, we use here the analytically obtained slope by Brilliantov~\cite{Brilliantov}.
We compare in figure~\ref{partHeLi} r.h.s. different analytical approximations for the Coulombic energy of a plama with the 
asymptotic by Brilliantov \cite{EbFoFi17} and the Debye energy.

We need a generalization of the KY theory which is appropriate for multi-charged ions as has been developed in \cite{EbFoFi17,Brilliantov}.
Let us consider a plasma with multi-charged ions denoted by indices $i$. According to Brilliantov, we do not permit the realization of high $k$-modes $k > k_i = (9  \piup^2 n_i)^{1/3}$, since the corresponding plasma modes cannot ne realized. Then, the integration should be restricted to lower modes and we get for the contribution of these species to the Coulomb energy and to the chemical potentials \cite{Fortov20,Brilliantov}
\bea
U_i^C = - k_{\rm B} T n_i Z_i^2 \sum_j Z_i^2 \frac{\kappa^2}{p_{ij}^2 - s_{ij}^2} m_{ij} .
\eea
Here, the matrix $m_{ij}$ is defined by
\bea 
m_{ii} = \bigg[ \frac{1}{p_{ii}} \arctan \bigg(\frac{k_{i}}{p_{ii}}\bigg) -  
\frac{1}{s_{ii}} \arctan \bigg(\frac{k_{i}}{s_{ii}}\bigg) \bigg], \\
m_{ij} =  \bigg[ \frac{1}{p_{ii}} - \frac{1}{s_{ii}} \bigg] \quad {\rm if} \quad i \ne j ,  \qquad k_i = (9 \piup^2 n_i)^{1/3}.
\eea
Note that this variant of the theory contains the free parameters $\alpha_{ij}$ and $k_i$ and may be adapted this way to 
specific tasks. The $p_{ij}$ and $s_{ij}$ are given by
\bea
 p_{ij} = \frac{\alpha_{ij}}{2} \bigg(\sqrt{1 - 2 \frac{\kappa}{\alpha_{ij}}} -\sqrt{1+ 2 \frac{\kappa}{\alpha_{ij}}}\bigg); \quad s = \frac{\alpha_{ij}}{2} \bigg(\sqrt{1 + 2 \frac{\kappa}{\alpha_{ij}}} +\sqrt{1 - 2 \frac{\kappa}{\alpha_{ij}}}\bigg).
\eea
We remember the relations \cite{Falkenhagen,Falkenhagen_a}
\bea
s_{ij} p_{ij} = \kappa^2; \qquad (s_{ij} + p_{ij}) (s_{ij} -p_{ij}) = \alpha_{ij} \sqrt{1 - 4 \kappa^2 / \alpha_{ij}^2} .
\eea
We mention that the method of mode restriction is of particular interest for the treatment of heavy ions in plasmas, since ions behave in plasmas in a completely different way than electrons and tend to form lattice-like structures, which leads to a dependence on $n_i^{1/3}$. However, we will not discuss this problem here in detail and refer to \cite{EbFoFi17,Fortov20,Stolzmann,Brilliantov,Eb16,Eb16_a}. 
Note, that in general, the transition to
restricted modes, i.e., to including $\arctan$-functions, may be relevant  only in case of strong ion-ion correlations

\section{Conclusions}
In this work we discuss the collaboration between the groups of Ihor R. Yukhnovskii in Lviv and G\"unter Kelbg in Rostock and analyze several approaches based on pair correlation functions and cluster expansion in the classical as well as in the quantum case. An essential role in this collaboration played
the exponential potential which has an easy Fourier transform and is in the quantum case equivalent to the Kelbg potential.
Finally, we discuss the progress in the statistical description of bound states of three particles as in  MgCl$_2$-solutions in the classical case and
in quantum helium plasmas. We presented new results about the effects of three-particle bound states, in particlular new expressions for the cluster integrals and the mass action functions of  ionic triple associates and helium atoms. 

We use here a chemical description derived from the fugacity expansion, having in mind
Onsager's view that we have some freedom in defining the chemical species. Like in writing the ledger for a 
company, we may write  some effects on a left-hand page of the lecture and some on a right-hand page. This has the same effect as far as a correct summation is observed. A chemical description is some kind of optimum arrangement between left-hand and right-hand (free and bound), since a minimization of the free energy is considered,
which, however, like in the case of a ledger, is strictly forbidden, and is any double counting of the terms. In the case of a ledger, this is against the law, while in the case of nature, it violates the laws of nature! Our cropping procedure is a direct consequence
of avoiding the double counting of terms. Not observing this, means to admit errors.

Our new results for the equation of state of hydrogen and helium plasmas are essential based on the lectures given by G\"unter Kelbg and this author on a seminar directed by Ihor R. Yukhnovskii in 1970 in the Institute on the Dragomanov street in Lviv.  This way we demonstrate here the key results of about 60 years of collaboration
between two groups working in Lviv and in Rostock, which was full of hard work connected as well as with many unforgettable  
warm personal meetings with Ihor Rafailovich and his colleagues. 

\section*{Acknowledgement}
We express our thanks to many colleagues for suggestions, providing additional material and encouragement, in particular 
to Gerd R\"opke. Further we express our thanks for a long collaboration with several colleagues during their time at Rostock University, such as David Blaschke, Michael Bonitz. Heinz Hoffmann, the late Dietrich Kremp, Wolf-Dietrich Kraeft, Heidi Reinholz,  and others. We gratefully remember 
in particular the collaboration with the late Hartmut Krienke which resulted in the most essential results for the classical case. A special sincere thank you goes to Myroslav Holovko from the Institute in Lviv, who was the permanent link between the two schools and held the contacts even in difficult times.

\ukrainianpart

\title{Статистична теорія систем заряджених частинок, включаючи потрійні зв'язані стани, та співпраця Львів--Росток}
\author{В. Ебелінг}
\address{
	Інститут фізики, Університет Гумбольдта, Берлін, Німеччина
}

\makeukrtitle

\begin{abstract} 
	Вшановуючи сторічний ювілей від дня народження Ігоря Р. Юхновського, ми аналізуємо нові досягнення в статистичній термодинаміці кулонівських систем. Основна ідея цієї роботи полягає в демонстрації того, що експоненціальний потенціал, використаний у перших статтях І.~Р.~Юхновського, є адекватною системою відліку для опису класичних та квантових систем заряджених частинок. Ми коротко обговорюємо співпрацю між групами Ігоря Юхновського у Львові та Гюнтера Кельба в Ростоку та аналізуємо деякі підходи, що грунтуються на парних кореляційних функціях та кластерних розвиненнях як у класичному, так і в квантовому випадку.
	Насамкінець, ми розповідаємо про прогрес у статистичному описі зв'язаних станів трьох частинок, таких як у плазмі гелію та в розчинах MgCl$_2$ у класичному випадку, та представляємо нові результати щодо впливу тричастинкових зв'язаних станів. Зокрема, ми наводимо нові вирази для кластерних інтегралів та констант рівноваги. у випадку атомів гелію та іонних потрійних асоціатів,
	а також для рівняння стану.
	
	\keywords 
	кореляційні функції, розклади за фугітивністю, константи хімічної рівноваги, утворення атомів 
\end{abstract}

\lastpage

\begin{thebibliography}{99}
\bibitem{Yukhn02} Yukhnovskii I., Selected Works. Physics,  Publishing House of Lviv Polytechnic National University, Lviv, 2002, (in Ukrainian).
\bibitem{Falkenhagen}
 Falkenhagen H., Theorie der Elektrolyte, Hirzel Leipzig, 1971, (in German).
%
\bibitem{Falkenhagen_a}
 Falkenhagen H., Ebeling W., In: Satyendranath Bose 70th birthday commemoration volume, Vol. 2, Prof.~S.~N.~Bose 70th Birthday Celebration Committee [1965-66], Calcutta, 1966.
\bibitem{Barthel}
 Barthel J., Krienke H., Kunz W., Physical Chemistry of Electrolyte Solutions: Modern Aspects, Topics in physical chemistry, Vol. 5, Springer, New York, 1998.
\bibitem{Bogoliubov}
 Bogoliubov N. N., Nonequilibrium Statistical Mechanics, 1939–1980, Statistical Mechanics, Vol. 5, Nauka, Moscow, 2006, (in Russian).
 %
 \bibitem{Bogoliubov_a}
 Bogoliubov N. N., Equilibrium Statistical Mechanics, 1945–1986, Statistical Mechanics, Vol. 6, Nauka, Moscow, 2006, (in Russian).
\bibitem{Friedman}
 Friedman H. L., Ionic Solution Theory: Based on Cluster Expansion Methods, Interscience Publishers, New~York, 1962.
\bibitem{GlaYuk52}
 Glauberman A. E., Yukhnovskii I. R., 
 Zh. Eksp. Teor. Fiz., 1952, {\bf 22}, 562--572, (in Russian). 
\bibitem{Yukhn54}
 Yukhnovskii I. R., 
 Zh. Eksp. Teor. Fiz., 1954, {\bf 27}, 690--698, (in Russian). 
%
\bibitem{Yukhn58}
 Iukhnovskii I. R., Sov. Phys. JETP, 1958, {\bf 7}, No. 2, 263--270.
\bibitem{Yukhn80}
 Yukhnovskii I. R., Holovko M. F., Statistical Theory of the Classical Equilibrium Systems, Naukova Dumka, Kiev, 1980, (in Russian).
\bibitem{Kelbg59}
 Kelbg G., Wiss. Z. U. Rostock MNR, 1959/60, {\bf 9}, 41. 
\bibitem{Kelbg59a}
 Kelbg G., In: Electrolytes,  Pesce B. (Ed.), Pergamon Press LTD, New York, London, 1962, 109.
\bibitem{Kelbg62}
 Kelbg G., Ann. Phys., 1962, {\bf 464}, 159--167, \doi{10.1002/andp.19624640307}, (in German). 
\bibitem{Kelbg62a} 
Kelbg G., Ann. Phys., 1963, {\bf 467}, 354--360, \doi{10.1002/andp.19634670703}, (in German).
 
 
\bibitem{Kelbg63}
 Kelbg G., Ann. Phys., 1963, {\bf 467}, 219--224, \doi{10.1002/andp.19634670308}, (in German).
 \bibitem{KelbgHoff64}
 Kelbg G., Hoffmann H. J.,  Ann. Phys., 1963, {\bf 469}, 310--318, \doi{10.1002/andp.19644690508}, (in German).
\bibitem{BoEbal22}
 Bonitz M.,  Ebeling W., Filinov A., Kraeft W. D., Redmer R., R\"opke G.,  	Contrib. Plasma Phys., 2023, {\bf 63}, e202300029, \doi{10.1002/ctpp.202300029}.
\bibitem{EbKe66}
Ebeling W., Kelbg G., Z. Phys. Chem., 1966, {\bf 233}, 209--230, \doi{10.1515/zpch-1966-23326}, (in German). 
\bibitem{EbKe66a}
Kelbg G., Ebeling W., Krienke H., Z. Phys. Chem., 1968, {\bf 238}, 76--88, \doi{10.1515/zpch-1968-23811}, (in German). 
\bibitem{Albehrendt}
 Albehrendt N., Bigun G.,  Juchnovski I. R., Kelbg G., Ann. Phys.
1970, {\bf 479}, 188--199,\\ \doi{10.1002/andp.19704790503}, (in German).
\bibitem{EbKr71}
 Ebeling W., Krienke H., Z. Phys. Chem., 1971, {\bf 248}, 274--276, \doi{10.1515/zpch-1971-24836}.
\bibitem{Czerwon}
Krienke H., Ebeling W., Czerwon H. J., Wiss. Z. Univ. Rostock MNR, 1975, {\bf 24}, 671--679.
\bibitem{EbKr79}
 Ebeling W., Krienke H., Rostock Physical Manuscripts, 1979, {\bf 4}, 55--66.
\bibitem{GoKr89}
 Golovko M. F., Krienke H., Mol. Phys., 1989, {\bf 68}, 967, \doi{10.1080/00268978900102671}.
\bibitem{Krienke2013}
Krienke H., Condens. Matter Phys., 2013, {\bf 16}, 43006, \doi{10.5488/CMP.16.43006}.
\bibitem{EbFeKr21}
 Ebeling W., Feistel R., Krienke H., 
 On statistical calculations of individual ionic activity coefficients of electrolytes and seawater. I. Basics - Draft 14 Apr 2019, 2019, \doi{10.13140/RG.2.2.18591.20640}.
 \bibitem{EbFeKr21a}
 Ebeling W., Feistel R., Krienke H., 
J. Mol. Liq., 2022, {\bf 346}, 117814, \doi{10.1016/j.molliq.2021.117814}.
\bibitem{EbMatrix24}
 Ebeling W., Contrib. Plasma Phys., 2024, {\bf 64}, e202300161, \doi{10.1002/ctpp.202300161}.
\bibitem{EbHoKuYu23}
Ebeling W., Holovko M., Kunz W., Yukhnovskii I., Condens. Matter Phys., 2023, {\bf 26}, No.~4, 47002,\\ \doi{10.5488/CMP.26.47002}.
\bibitem{EFKS77}
 Conway B. E., Outhwaite C. W., Stell G., Ebeling W., Valleau J. P., Adams D. J., Friedman H. L., Symons M. C. R., Hertz H. G., Newman K. E., et. al., Faraday Discuss. Chem. Soc., 1977, {\bf 64}, 69--94, \doi{10.1039/DC9776400069}.
%
\bibitem{EFKS77_a}
Kay R. L., Justice J. C., Ebeling W., Fuoss R. M., Glueckauf E., Pethybridge A., Friedman H. L., Barthel~J., Wolynes P. G., Krumgal{'}z B., Stead K., Rosseinsky D. R., Spiro M., Rossotti F. J. C., Symons M. C. R., Irish~D.~E., Miller D. G., Paterson R., Hertz H. G., Griffiths T. R.,  Faraday Discuss. Chem. Soc., 1977, {\bf 64}, 322--353, \doi{10.1039/DC9776400322}.
\bibitem{EbFeSa79}
Ebeling W., Feistel R., S\"andig R., J. Solution Chem., 1979, {\bf 8},  53--82, \doi{10.1007/BF00646809}.
\bibitem{EbKr2023}
Ebeling W., Krienke H., Condens. Matter Phys., 2023, {\bf  26},
No.~2, 23602, \doi{10.5488/CMP.26.23602}.
\bibitem{EbHoKe67}
Ebeling W., Hoffmann H., Kelbg G., 	Contrib. Plasma Phys., 1967, {\bf 7},  233, \doi{10.1002/ctpp.19670070307}.
\bibitem{EbHoKe67a}
Hoffmann H. J., Ebeling W., Physica, 1968, {\bf 39},  593, \doi{10.1016/0031-8914(68)90034-7}. 
\bibitem{EbHoKe67b}
Ebeling W., Physica, 1968, {\bf 40},  290, \doi{10.1016/0031-8914(68)90025-6}.
\bibitem{KelbgEb70}
Kelbg G., Ebeling W., Preprint of the Institute for Theoretical Physics, ITF–70–93P, Kiev, 1970, (in Russian).
\bibitem{KelbgEb70a}
Kelbg G., Ebeling W., Preprint of the Institute for Theoretical Physics, ITF–70–94P, Kiev, 1970, (in Russian).

\bibitem{Kramers}
 Kramers H. A., Collected Scientific Papers, Amsterdam, 1956.
\bibitem{Hellmann}
 Hellmann H., 
 Acta Physicochim. URSS, 1934/1935, {\bf 1}, 333--353, (in German).
\bibitem{Hellmann_a}
Hellmann H., Acta Physicochim. URSS, 1936,  {\bf 4}, 225--244, (in German).
\bibitem{Hellmann_b}
Andrae D. (Ed.), Hans Hellmann: Einf\"uhrung in die Quantenchemie, Springer Spektrum Berlin, Heidelberg, 2015, \doi{10.1007/978-3-662-45967-6}.
\bibitem{BaEb71}
 Bartsch G. P., Ebeling W., Contrib. Plasma Phys., 1971, {\bf 11}, 393, \doi{10.1002/ctpp.19710110505}.
\bibitem{Kelbg72}
Kelbg G., In: Ergebnisse der Plasmaphysik und Gaselektronik, Vol. III, Rompe R., Steenbecs (Eds.), Akademie-Verlag, Berlin, 1972, 363--434, (in German).
\bibitem{FriedmanEb79}
Friedman H. L., Ebeling W., Rostock Physical Manuscripts, 1979, {\bf 4}, 33.
\bibitem{EKK76}
 Ebeling W., Kraeft W. D., Kremp D., Theory of Bound States and Ionization Equilibrium in Plasmas and Solids, Akademie-Verlag, Berlin, 1976.
\bibitem{Balescu}
 Balescu R., Equilibrium and Non-Equilibrium Statistical Mechanics, Wiley, New York, 1975. 
\bibitem{Rogers}
Rogers F. J., Phys. Rev. A, 1974, {\bf 10}, 2441, \doi{10.1103/PhysRevA.10.2441}.
\bibitem{Rogers_a}
Rogers F. J.,  Phys. Rev. A, 1981, {\bf 24}, 1531, \doi{10.1103/PhysRevA.24.1531}. 
\bibitem{Rogers79}
Rogers F. J., Phys. Rev. A, 1979, {\bf 19}, 375, \doi{10.1103/PhysRevA.19.375}.
\bibitem{Rogers79_b}
Rogers F. J., Contrib. Plasma Phys., 2001, {\bf 41}, 179, \doi{10.1002/1521-3986(200103)41:2/3<179::AID-CTPP179>3.0.CO;2-H}.
\bibitem{Rogers86}
Rogers F. J., 	Astrophys. J., 1986, {\bf 310}, 723, \doi{10.1086/164725}.
\bibitem{Rogers01}
Rogers F. J., DeWitt H. E., Contrib. Plasma Phys., 2003, {\bf 43}, 355, \doi{10.1002/ctpp.200310045}.
\bibitem{Klim75}
Klimontovich Yu. L., Kinetic Theory of Nonideal Gases and Nonideal Plasmas,  Academic Press, New York, 1982.
\bibitem{KKER86}
Kremp D., Kraeft W. D, Ebeling W., R\"opke G., Quantum Statistics of Charged Particle Systems, Plenum Press, New York, 1986.
\bibitem{EbFoFi17}
Ebeling W., Fortov V. E., Filinov V.,  Quantum Statistics of Dense Gases and Nonideal Plasmas, Springer Nature, Cham, Switzerland, 2017.
\bibitem{Fortov20}
Fortov V. E., Filinov V. S., Larkin A. S., Ebeling W., Statistical Physics of Dense Gases and Nonideal Plasmas, FizMatLit, Moscow, 2020, (in Russian).
\bibitem{Alastuey95}
 Alastuey A., Cornu F., Perez A., Phys. Rev. E, 1995, {\bf 51}, 1725, \doi{10.1103/PhysRevE.51.1725}.
\bibitem{Alastuey08}
Alastuey A., Ballenegger V., Cornu F., Martin Ph. A., J. Stat. Phys., 2008, {\bf 130}, 1119, \doi{10.1007/s10955-007-9464-0}.
\bibitem{Alastuey08_a}
Alastuey A., Ballenegger V., Contrib. Plasma Phys., 2010,  {\bf 50}, 46--53, \doi{10.1002/ctpp.201010011}.
\bibitem{KrScKr05}
Kremp D., Schlanges M., Kraeft W. D., Quantum Statistics of Nonideal Plasmas, Springer, Berlin, 2005.
\bibitem{NIST}
Kramida A., Ralchenko Y., Reader J.,  NIST ASD Team, NIST Atomic Spectra Database, NIST Standard Reference Database 78 (ver. 5.10), 2022, \doi{10.18434/T4W30F}.
\bibitem{BonitzPIMC}
Filinov A.,  Bonitz M., Phys. Rev. E, 2023, {\bf 108}, 055212, \doi{10.1103/PhysRevE.108.055212}.
\bibitem{BonitzPIMC_a}
Bonitz M., Vorberger J.,  Bethkenhagen M., B\"ohme M. P., Ceperley~D.~M.,  Filinov~A., Gawne T., Graziani F., Gregori G., Hamann P., et. al., Phys. Plasmas, 2024, {\bf 31}, 110501, \doi{10.1063/5.0219405}.
\bibitem{Unsold}
Uns\"old A., Physik der Sternatmosph\"aren, Springer, Berlin, 1956, (in German).
\bibitem{Stolzmann}
Stolzmann W., Ebeling W., Phys. Lett. A, 1998, {\bf 249}, 242, \doi{10.1016/S0375-9601(98)00659-8}.
\bibitem{Brilliantov}
Brilliantov N., Contrib. Plasma Phys., 1998, {\bf 38}, 489, \doi{10.1002/ctpp.2150380403}.
\bibitem{Eb16}
 Ebeling W., Contrib. Plasma Phys., 2016, {\bf 56}, 163, \doi{10.1002/ctpp.201500118}. 
%
\bibitem{Eb16_a}
 Ebeling W., Pseudopotentials and Pair Correlation Function for Nonideal Plasma, Proc. Conf., Almaty.
\bibitem{EbKrRo12}
Ebeling W., Kraeft W. D., R\"opke G., Ann. Phys., 2012, {\bf 524}, 311, \doi{10.1002/andp.201100331}.
\bibitem{EbKrRo12_a}
Ebeling W., Kraeft W. D., R\"opke G., Contrib. Plasma Phys., 2012, {\bf 52}, 7, \doi{10.1002/ctpp.201100056}.
\bibitem{EbRoRe21}
Ebeling W., Reinholz H., R\"opke G., Eur. Phys. J. Spec. Top., 2020, {\bf 229}, 3403, \doi{10.1140/epjst/e2020-000066-6}. 
\bibitem{EbRoRe21_a}
Ebeling W., Reinholz H., R\"opke G., Contrib. Plasma Phys., 2021, {\bf 61}, 10, \doi{10.1002/ctpp.202100085}.
\bibitem{EbRoMDPI}
Ebeling W., R\"opke G.,  Plasma, 2023, {\bf 6}, 1--26, \doi{10.3390/plasma6010001}.
\bibitem{EbRoCPP24}
Ebeling W., R\"opke G., 
 (unpublished).
\bibitem{Cardona10}
 Cardona O., Mart\'inez-Arroyo M., L\'opez-Castillo M. A., Astrophys. J., 2010, {\bf 711}, 239, \doi{10.1088/0004-637X/711/1/239}.

\end{thebibliography}
\end{document}